\documentclass[useAMS, usenatbib, longauth]{aa}

\usepackage{graphicx}
\usepackage{txfonts}
\usepackage{lscape}
\usepackage{amssymb,amsmath}
\usepackage{rotating}
\usepackage{xcolor}
\usepackage{hyperref}
\usepackage{comment}
\usepackage{adjustbox}
\usepackage{multirow}
\usepackage{subfigure}
\usepackage{soul}
\usepackage{placeins}
\usepackage{cuted}
\usepackage{tabularx}
\usepackage{stfloats}
\usepackage{ulem}

\hypersetup{
    colorlinks=true,
    linkcolor=blue,
    filecolor=blue,      
    urlcolor=blue,
    citecolor=blue,
    pdftitle={Constraining Lens Masses in Moderately to Highly Magnified Microlensing Events from Gaia},
    pdfpagemode=FullScreen, 
    }
\def\msun{M_\odot}

\def\I0st{{I_{\mathrm 0}^{\mathrm{st}}}}
\def\V0{V_{\mathrm 0}}

\def\t0{t_{\mathrm 0}}

\def\u0{u_{\mathrm 0}}

\def\thetaE{\theta_{\mathrm{E}}}

\newcommand{\orcid}[1]{\protect\href{https://orcid.org/#1}{\protect\includegraphics[width=8pt]{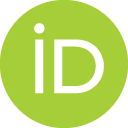}}}

\begin{document} 

  \title{Constraining lens masses in moderately
to highly magnified microlensing events from Gaia}

\titlerunning{Constraining lens masses in moderately
to highly magnified microlensing events from Gaia}
\authorrunning{Uliana Pylypenko et al.}

   \author{
   U. Pylypenko \orcid{0009-0002-7560-1903} \inst{\ref{OA}}\fnmsep\thanks{\email{uliana.pylypenko@gmail.com}},
   {\L}. Wyrzykowski \orcid{0000-0002-9658-6151} \inst{\ref{OA},\ref{NCBJ}},
P. J. Miko{\l}ajczyk \orcid{0000-0001-8916-8050} \inst{\ref{OA},\ref{NCBJ},\ref{UWr}},
K. Kotysz \orcid{0000-0003-4960-7463} \inst{\ref{OA},\ref{UWr}},
P. Zieli{\'n}ski \orcid{0000-0001-6434-9429} \inst{\ref{UMK}},
N. Ihanec \inst{\ref{OA}},
M. Wicker \orcid{0009-0006-7642-0943} \inst{\ref{OA}}, 
M. Ratajczak \orcid{0000-0002-3218-2684} \inst{\ref{OA}},
M. Sitek \orcid{0000-0002-1568-4551} \inst{\ref{OA}},
K. Howil \orcid{0000-0002-4085-935X} \inst{\ref{OA}, \ref{mii}},
M. Jab{\l}o{\'n}ska \orcid{0000-0001-6962-4979} \inst{\ref{OA},\ref{anu}},
Z. Kaczmarek \orcid{0009-0007-4089-5012} \inst{\ref{heidelberg}},
K. Kruszy{\'n}ska \orcid{0000-0002-2729-5369} \inst{\ref{OA},\ref{lco}},
A. Udalski \orcid{0000-0001-5207-5619} \inst{\ref{OA}},
G. Damljanovic \orcid{0000-0002-6710-6868} \inst{\ref{Belgrade}},
M. Stojanovic \orcid{0000-0002-4105-7113} \inst{\ref{Belgrade}},
M. D. Jovanovic \orcid{0000-0003-4298-3247} \inst{\ref{Belgrade}},
T. Kvernadze \orcid{0000-0001-9947-4983} \inst{\ref{Abatsumani}},
O. Kvaratskhelia \inst{\ref{Abatsumani}},
M. {\.Z}ejmo \orcid{0000-0001-5836-9503} \inst{\ref{Zielona Gora}},
S. M. Brincat \orcid{0000-0002-9205-5329} \inst{\ref{Flarestar}},
J. K. T. Qvam \inst{\ref{Horten}},
T. G\"uver \orcid{0000-0002-3531-9842} \inst{\ref{Istanbul}},
E. Bachelet \orcid{0000-0002-6578-5078} \inst{\ref{IPAC}},
K. A. Rybicki \orcid{0000-0002-9326-9329} \inst{\ref{WIS}},
A. Garofalo \orcid{0000-0002-5907-0375} \inst{\ref{Bologna}},
J. Zdanavicius \inst{\ref{Vilnius}},
E. Pakstiene \inst{\ref{Vilnius}},
S. Zola \orcid{0000-0003-3609-382X} \inst{\ref{OAUJ}},
S. Kurowski \orcid{0000-0002-1557-0343} \inst{\ref{OAUJ}},
D. E. Reichart \orcid{0000-0002-5060-3673} \inst{\ref{skynet1}},
J. W. Davidson Jr. \orcid{0009-0007-1284-7240} \inst{\ref{skynet2}},
U. Burgaz \orcid{0000-0003-0126-3999} \inst{\ref{tcd}},
J. P. Rivet \orcid{0000-0002-0289-5851} \inst{\ref{France}}, 
M. Jelinek \orcid{0000-0003-3922-7416} \inst{\ref{Ondrejov}},
A. Popowicz \inst{\ref{Gliwice}},
H. H. Esenoglu \orcid{0000-0003-3531-7510} \inst{\ref{Istanbul}},
E. Sonbas \orcid{0000-0002-6909-192X} \inst{\ref{Adiyaman},\ref{Washington}},
J. M. Carrasco \orcid{0000-0002-3029-5853} \inst{\ref{iccub},\ref{fqa},\ref{ieec}},
S. Awiphan \orcid{0000-0003-3251-3583}\inst{\ref{Thailand}},
O. Tasuya \inst{\ref{Thailand}},
V. Godunova \orcid{0000-0001-7668-7994} \inst{\ref{icamer}},
A. Simon \orcid{0000-0003-0404-5559} \inst{\ref{Glushkova 4},\ref{MAN}},
A. Fukui \orcid{0000-0002-4909-5763} \inst{\ref{Komaba}},
C. Galdies \orcid{0000-0002-8908-0785} \inst{\ref{Malta},\ref{Znith}},
K. B\k{a}kowska \orcid{0000-0003-1034-1557} \inst{\ref{UMK}},
P. Hofbauer \inst{\ref{UMK}},
A. Gurgul \orcid{0000-0002-9441-0195} \inst{\ref{UMK}},
B. Joachimczyk \orcid{0009-0005-1710-6754} \inst{\ref{UMK}},
M. Dominik \inst{\ref{StAndrews}},
F. Cusano \inst{\ref{Bologna}},
I. Ilyin \inst{\ref{Leibniz}}\\
and \\
Y. Tsapras \orcid{0000-0001-8411-351X} \inst{\ref{heidelberg}},
R. A. Street \inst{\ref{lco}},
M. Hundertmark \orcid{0000-0003-0961-5231} \inst{\ref{heidelberg}},
V. Bozza \inst{\ref{Salerno},\ref{Napoli}},
P. Rota \inst{\ref{Salerno},\ref{Napoli}},
A. Cassan \inst{\ref{Sorbonne}},
J. Wambsganss \inst{\ref{heidelberg}},
R. Figuera Jaimes \inst{\ref{Chile1},\ref{Chile2}}\\
(The OMEGA Key Project)
}

\institute{Astronomical Observatory, University of Warsaw, Al. Ujazdowskie~4, 00-478 Warsaw, Poland
    \label{OA}
    \and
    Astrophysics Division, National Centre for Nuclear Research, Pasteura 7, 02-093, Warsaw, Poland 
    \label{NCBJ}
    \and
    Astronomical Institute, University of Wroc{\l}aw, ul. Miko{\l}aja Kopernika 11, 51-622 Wroc{\l}aw, Poland
    \label{UWr}
    \and
    Institute of Astronomy, Faculty of Physics, Astronomy and Informatics, Nicolaus Copernicus University in Toru{\'n}, Grudzi\k{a}dzka 5, 87-100 Toru{\'n}, Poland
    \label{UMK}
    \and
    Faculty of Mathematics and Computer Science, Jagiellonian University, {\L}ojasiewicza 6, 30-348 Krak{\'o}w, Poland
    \label{mii}
    \and
    Research School of Astronomy and Astrophysics, Australian National University, Mount Stromlo Observatory, Cotter Road Weston Creek, ACT 2611, Australia
    \label{anu}
    \and
    Zentrum f{\"u}r Astronomie der Universit{\"a}t Heidelberg, Astronomisches Rechen-Institut, M{\"o}nchhofstr. 12-14, 69120 Heidelberg, Germany
    \label{heidelberg}
    \and
    Las Cumbres Observatory Global Telescope Network, 6740 Cortona Drive, suite 102, Goleta, CA 93117, USA
    \label{lco}
    \and
    Astronomical Observatory, Volgina 7, 11060 Belgrade, Serbia
    \label{Belgrade}
    \and
    E.Kharadze Georgian National Astrophysical Observatory, 0301 Abastumani, Georgia
    \label{Abatsumani}
    \and
    Janusz Gil Institute of Astronomy, University of Zielona G{\'o}ra, Lubuska 2, PL-65-265 Zielona G{\'o}ra, Poland
    \label{Zielona Gora}
    \and
    Flarestar Obsevatory, Fl.5 Ent.B, Silver Jubilee Apt, George Tayar Street, San Gwann, SGN 3160, Malta
    \label{Flarestar}
    \and
    Horten Videregaende Skole, Strandpromenaden 33, 3183 Horten, Norway
    \label{Horten}
    \and
    Istanbul University, Faculty of Science, Department of Astronomy and Spaces Sciences, 34119, Istanbul Türkiye, Istanbul University Observatory Research and Application Center, Istanbul University 34119, Istanbul T\"{u}rkiye
    \label{Istanbul}
    \and
    Department of Physics, Ad{\i}yaman University, Ad{\i}yaman 02040, T\"{u}rkiye
    \label{Adiyaman}
    \and
    Department of Physics, The George Washington University, Washington, DC 20052, USA
    \label{Washington}
    \and
    IPAC, Mail Code 100-22, Caltech, 1200 E. California Blvd., Pasadena, CA 91125, USA
    \label{IPAC}
    \and
    Department of Particle Physics and Astrophysics, Weizmann Institute of Science, Rehovot 76100, Israel
    \label{WIS}
    \and
    INAF-Osservatorio di Astrofisica e Scienza dello Spazio, Via Gobetti 93/3, I-40129 Bologna, Italy
    \label{Bologna}
    \and
    Department of Physics and Astronomy, University of Sheffield, Sheffield, S3 7RH, UK
    \label{Sheffield}
    \and
    Institute of Theoretical Physics and Astronomy, Vilnius University, Saulėtekio Av. 3, 10257 Vilnius, Lithuania
    \label{Vilnius}
    \and
    Astronomical Observatory, Jagiellonian University, ul. Orla 171, PL-30-244 Kraków, Poland
    \label{OAUJ}
    \and
    University of North Carolina at Chapel Hill, Chapel Hill, North Carolina, NC 27599, USA
    \label{skynet1}
    \and
    Department of Astronomy, University of Virginia, 530 McCormick Rd., Charlottesville, VA, 22904, USA
    \label{skynet2}
    \and
    School of Physics, Trinity College Dublin, College Green, Dublin 2, Ireland
    \label{tcd}
    \and
    Universit{\'e} {\^C}ote d'Azur, Observatoire de la {\^C}ote d'Azur, CNRS, Laboratoire Lagrange, France
    \label{France}
    \and
    Astronomical Institute of the Academy of Sciences of the Czech Republic (ASU CAS), 25165 Ondrejov, Czech Republic
    \label{Ondrejov}
    \and
    Faculty of Automatic Control, Electronics and Computer Science, Silesian University of Technology, Akademicka 16, 44-100 Gliwice, Poland
    \label{Gliwice}
    \and
    Institut de Ci\`encies del Cosmos (ICCUB), Universitat de Barcelona (UB), Mart\'{i} i Franqu\`es 1, E-08028 Barcelona, Spain
    \label{iccub}
    \and
    Departament de F\'isica Qu\`antica i Astrof\'{i}sica (FQA), Universitat de Barcelona (UB), Mart\'{i} i Franqu\`es 1, E-08028 Barcelona, Spain
    \label{fqa}
    \and
    Institut d'Estudis Espacials de Catalunya (IEEC), Esteve Terradas, 1, Edifici RDIT, Campus PMT-UPC, 08860 Castelldefels (Barcelona), Spain
    \label{ieec}
    \and
    National Astronomical Research Institute of Thailand (Public Organization), 260 Moo 4, Donkaew, Mae Rim, Chiang Mai 50180, Thailand
    \label{Thailand}    
    \and
    ICAMER Observatory of National Academy of Sciences of Ukraine, 27 Acad. Zabolotnoho str., Kyiv, 03143 Ukraine
    \label{icamer}
    \and
    Astronomy and Space Physics Department, Taras Shevchenko National University of Kyiv, 4 Glushkova ave., Kyiv, 03022 Ukraine
    \label{Glushkova 4}
    \and
    National Center "Junior Academy of Sciences of Ukraine", 38-44 Dehtiarivska St., Kyiv, 04119 Ukraine
    \label{MAN}
    \and
    Komaba Institute for Science, The University of Tokyo, 3-8-1 Komaba, Meguro, Tokyo 153-8902, Japan
    \label{Komaba} 
    \and
    Institute of Earth Systems, University of Malta
    \label{Malta}
    \and
    Znith Astronomy Observatory, Malta
    \label{Znith}
    \and
    University of St Andrews, Centre for Exoplanet Science, SUPA School of Physics \& Astronomy, North Haugh, St Andrews, KY16 9SS, United Kingdom
    \label{StAndrews}
    \and
    Dipartimento di Fisica "E.R. Caianiello", Universit{\`a} di Salerno, Via Giovanni Paolo II 132, I-84084 Fisciano, Italy
    \label{Salerno}
    \and
    Istituto Nazionale di Fisica Nucleare, Sezione di Napoli, Via Cintia, I-80126 Napoli, Italy
    \label{Napoli}
    \and
    Institut d'Astrophysique de Paris, Sorbonne Universit\'e, CNRS, UMR 7095, 98 bis bd Arago, F-75014 Paris, France
    \label{Sorbonne}
    \and
    Millennium Institute of Astrophysics MAS, Nuncio Monsenor Sotero Sanz 100, Of. 104, Providencia, Santiago, Chile
    \label{Chile1}
    \and
    Instituto de Astrof\'isica, Facultad de F\'isica, Pontificia Universidad Cat\'olica de Chile, Av. Vicu\~na Mackenna 4860, 7820436 Macul, Santiago, Chile
    \label{Chile2}
    \and
    Leibniz-Institut für Astrophysik Potsdam (AIP), An der Sternwarte 16, 14482 Potsdam, Germany
    \label{Leibniz}
}

\date{Received 1 April 2025 / Accepted 4 September 2025}

\abstract
{Microlensing events provide a unique way to detect and measure the masses of isolated, non-luminous objects, particularly dark stellar remnants. Under certain conditions, it is possible to measure the mass of these objects using photometry alone, specifically when a microlensing light curve displays a finite source (FS) effect. This effect generally occurs in highly magnified light curves, i.e. when the source and the lens are very well aligned.}
{In this study, we analyse Gaia Alerts and Gaia Data Release 3 datasets, identifying four moderate-to-high-magnification microlensing events without a discernible FS effect. The absence of this effect suggests a large Einstein radius, implying substantial lens masses.
}
{In each event, we constrained the FS effect, and therefore established lower limits for the angular Einstein radius and lens mass. Additionally, we used the \texttt{DarkLensCode} software to obtain the mass, distance, and brightness distribution for the lens based on the Galactic model.}
{Our analysis established lower mass limits of $\sim 0.7$ $\msun$ for one lens and $\sim 0.3-0.5$ $\msun$ for two others. A \texttt{DarkLensCode} analysis supports these findings, estimating lens masses in the range of $\sim 0.42-1.70$ $\msun$ and dark lens probabilities exceeding 80\%. These results strongly indicate that the lenses are stellar remnants, such as white dwarfs or neutron stars.
}
{While further investigations are required to confirm the nature of these lenses, we demonstrate a straightforward yet effective approach to identifying stellar remnant candidates.
}


   \maketitle

\section{Introduction} \label{sec:introduction}

Microlensing is a powerful technique, allowing for the detection of isolated objects regardless of their luminosity. This makes it particularly valuable for studying dark, isolated stellar remnants in our Galaxy. The ability to identify more isolated black holes, neutron stars, and white dwarfs provides a unique opportunity to explore the late stages of stellar evolution and the nature of supernovae.

Despite its potential, microlensing remains a rare phenomenon. Detecting such events requires large-scale sky surveys that monitor millions of stars, such as Gaia \citep{Gaia}. Operating until January 2025, Gaia scanned the entire sky approximately once per month, leading to the discovery of numerous microlensing events \citep{Wyrzykowski2022, Hodgkin2021}. These observations have already resulted in the identification of several stellar remnant candidates (e.g. \cite{Jablonska2022, Kruszynska2022, Kruszynska2024, Howil2024}).

Determining the mass of a microlensing object requires two key parameters: the microlens parallax, $\pi_\mathrm{E}$, and the angular Einstein radius, $\theta_\mathrm{E}$. Once these parameters are obtained, the lens mass, $M_\mathrm{L}$, can be calculated using the relation \citep{Gould2000b}

\begin{equation}
    M_\mathrm{L} = \frac{\theta_\mathrm{E}}{\kappa \pi_\mathrm{E}},
    \label{eq:mass}
\end{equation}

where $\kappa = 8.144 \frac{\mathrm{mas}}{\msun}$. If the distance to the source is known, the distance to the lens, $D_\mathrm{L}$, can be determined from the equation

\begin{equation}
    D_\mathrm{L} = \biggl(\frac{\theta_\mathrm{E} \pi_\mathrm{E}}{1 \mathrm{AU}} + \frac{1}{D_\mathrm{S}}\biggr)^{-1}
    \label{eq:distance}
.\end{equation}

Microlensing events involving stellar remnants typically have long timescales, allowing for the measurement of the annual microlens parallax from the light curve. This effect arises due to Earth's motion around the Sun and appears as characteristic deviations, such as asymmetries, in the light curve \citep{Gould2000b}. This technique has been successfully applied to numerous microlensing events (e.g. \cite{Kruszynska2022, Howil2024}).

Measuring the Einstein radius is generally more challenging. One theoretically universal method applicable to any microlensing event involves detecting the astrometric counterpart -- an effect known as astrometric microlensing. The astrometric deviation, $\delta$, is directly proportional to the angular Einstein radius and reaches its maximum value at: $\delta_\mathrm{max} \approx 0.354 \theta_\mathrm{E}$ \citep{Dominik2000}. Since these deviations are typically of the order of a milliarcsecond or smaller, they are challenging to measure. This has been achieved twice using the Hubble Space Telescope (HST). The lens in one of the events, OGLE-2011-BLG-0462, is an isolated black hole with a mass of  $\sim 8$ $\msun$ (\cite{Sahu2022, Lam2022, Mroz2022, Lam2024}), while in the other event it is a well-known white dwarf LAWD 37 (\cite{McGill2023}). Apart from HST, Gaia has sufficient astrometric precision to detect astrometric microlensing. While these astrometric data are not yet publicly available, they are expected to provide numerous lens mass measurements in the future \citep{Rybicki2018}.

Another approach is the interferometric microlensing. During the microlensing event, where the lens is an isolated object, two source images with different brightness are produced. In this technique, the images are directly detected using interferometry. However, this is challenging in particular due to the small angular separation between the images. Therefore, this method has been successfully implemented only three times: in \cite{Dong2019}, \cite{Cassan2022}, and \cite{Mroz2025}.

A more straightforward way to measure the angular Einstein radius is by determining the source-lens relative proper motion, $\mu_\mathrm{rel}$. Then it can be obtained as $\theta_\mathrm{E} = \mu_\mathrm{rel} t_\mathrm{E}$, where $t_\mathrm{E}$ is a timescale of the event. This has been done several times; for example, in \cite{Bennet2024}, \cite{Retskini2024}, \cite{Bhattacharya2021}, and \cite{Vandorou2020}. Nevertheless, measuring the relative proper motion still requires advanced techniques and is only feasible if the lens is luminous, making it unsuitable for dark stellar remnants.

A less common method involves detecting the xallarap effect, which occurs when a source star is a binary system and its orbital motion influences the light curve. Despite the abundance of binary stars, this effect is expected to be present in just $\sim20$\% of the microlensing events, as a source system must have an optimal period with respect to the timescale of the event \citep{Poindexter2005, Hu2024}. Einstein radius measurement or estimations with xallarap were performed in, for example, \cite{Hu2024} and \cite{Smith2002}. 

Finally, the finite source (FS) effect, the primary focus of this study, requires specific conditions to be detected, but has been extensively used to measure the Einstein radius. It occurs when the projected separation between a source and a caustic is comparable to the source size \citep{Gould1994}. The magnitude of the FS effect is determined by the angular source radius in units of the angular Einstein radius of the event, typically denoted by $\rho$. By measuring this parameter and the angular source radius, the angular Einstein radius can be determined.

The FS effect is frequently detected in binary lens events. Most often such lenses are planetary systems (e.g. \cite{Wu2024, Shin2022, Yee2021}) or binary stars (e.g. \cite{Rota2024, Jung2015}). Masses of single lenses are also commonly measured. Most of them are single main-sequence stars (e.g. \cite{Rybicki2022, Zang2020, Yee2015Spitzer}). However, free-floating planets (e.g. \cite{Koshimoto2023, Mroz2020}) and isolated brown dwarfs (e.g. \cite{Shvartzvald2019}) have also been detected.

For the FS effect to be robust, there should be a moment during the event when the source-lens separation is of the order of the source size, and the event has to be actually observed during this period \citep{Gould1994}. In single-lens events, this typically requires a precise source-lens alignment, resulting in moderate-to-high magnification events. However, some events with significantly magnified curves do not exhibit the FS effect. Two primary reasons for this include a small source size and a large Einstein radius. The latter possibility is particularly promising in the context of searching for massive lenses such as black holes and other stellar remnants, as it may imply high mass. In cases in which there is no clear detection of the FS effect, it is not possible to measure $\rho$ directly, but it is possible to obtain its upper limit, $\rho_\mathrm{lim}$, and consequently a lower limit of the angular Einstein radius, $\theta_\mathrm{E,lim}$, which may be sufficient to infer the lens nature if combined with a lens light analysis.

Similar approaches have been adopted in several studies. \cite{Smith2002} determined $\rho_\mathrm{lim}$ and $\theta_\mathrm{E,lim}$, but it is not a strong constraint in this case, as the angular Einstein radius limit is quite small,  likely attributable to the relatively low magnification of the light curve ($A_\mathrm{max}\sim 10$). \cite{Shvartzvald2014}, \cite{Zang2018_FS}, and \cite{Bachelet2022} focused on binary lens events with relatively distant source-caustic passages and also derived $\rho_\mathrm{lim}$ using similar methods. The first two additionally estimated the angular Einstein radius limits (both <100 $\mu$as) and used Bayesian analysis for lens properties, while \cite{Bachelet2022} employed Markov chain Monte Carlo (MCMC) methods to refine constraints. In these cases, determining $\rho_\mathrm{lim}$ effectively extracts additional information from the light curve, leading to more meaningful constraints on lens properties. Our goal is to apply this technique more systematically.

In this study, we analyse four medium-to-high magnification microlensing events from Gaia that show no apparent FS effect and assess whether their lenses could be stellar remnant candidates. We constrain the angular Einstein radius and the lens mass using the distribution of the $\rho$ parameter, quantifying the FS effect. This constraint is based solely on photometric data, making it a simple yet effective method that can be applied to a broader range of events.

The structure of this paper is as follows. In Section \ref{sec:data}, we present the dataset and describe the cleaning process applied to ensure data quality. Section \ref{sec:method} outlines our methodology for analysing the microlensing events. The results for each event are detailed in Section \ref{sec:results}. We provide a broader discussion of our findings in Section \ref{sec:discussion} and summarise our conclusions in Section \ref{sec:conclusions}.

\section{Data and event selection} \label{sec:data}

\begin{table*}[h]
\small
\centering
\caption{Astrometric parameters for the sources from GDR3.}
\label{tab:astrometry}
\begin{tabular}{c|c|c|c|c}
\hline
  Parameter	&	Gaia21efs	&	Gaia21azb	&	GaiaDR3-ULENS-067	&	Gaia21dpb	\\
\hline
\hline

GDR3 ID & 1861870251275264896 & 4295398373281907200 & 4042928139682133120 & 1972242492638284160 \\
$\alpha$ [J2016.0]	&	20:29:41.89	&	19:22:33.35	&	18:01:09.17	&	21:25:29.89	\\
$\delta$ [J2016.0]	&	31:17:42.90	&	06:31:07.90	&	-32:33:27.72	&	46:25:08.83	\\
$\varpi$ [mas]	&	0.12 $\pm$ 0.03	&	0.12 $\pm$0.16	&	-0.03 $\pm$ 0.20	&	-0.47 $\pm$ 0.27	\\
$\mu_{\alpha}$ [mas yr$^{-1}$]	&	-2.83 $\pm$ 0.02	&	-2.80 $\pm$ 0.16	&	-2.01 $\pm$ 0.21	&	-2.91 $\pm$ 0.27	\\
$\mu_{\delta}$ [mas yr$^{-1}$]	&	-4.94$\pm$0.03	&	-6.68$\pm$0.17	&	-5.03 $\pm$ 0.13	&	-4.88 $\pm$ 0.32	\\
$\sigma_{\mu}$	&	-0.19	&	0.49	&	0.32	&	-0.02	\\
RUWE	&	0.97	&	1.02	&	2.35	&	1.02	\\
\hline
\end{tabular}

\tablefoot{Columns: GDR3 ID, co-ordinates ($\alpha$, $\delta$), parallax ($\varpi$), proper motions in directions of ($\alpha$, $\delta$) and their correlation ($\mu_{\alpha}$, $\mu_{\delta}$, $\sigma_{\mu}$), and re-normalised unit weight error (RUWE).}
\end{table*}

We focus on four microlensing events identified in Gaia data: Gaia21efs (ZTF21abqhbuh), Gaia21azb (ASASSN-21ht), GaiaDR3-ULENS-067 (OGLE-2015-BLG-0149), and Gaia21dpb (ZTF21abtulxp). In this section, we describe the photometric data collected for these events (Gaia photometry in Section \ref{sec:gaia-phot} and ground-based observations in Section \ref{sec:other-phot}), as well as the process of data cleaning (Section \ref{sec:data cleaning}). Additionally, we describe spectroscopic follow-up for Gaia21efs in Section \ref{sec:spectroscopy}.

 The events were selected based on the following criteria. First, the amplitude of the event's light curve had to be higher than 3$^m$. Second, the light curve had to be well covered, particularly at the peak. Third, the initial microlensing model (point source point lens model with parallax, more details in Section \ref{sec:phot-model}) had to satisfy the conditions $u_0 \lesssim 0.1$ and $\rho \lesssim u_\mathrm{0}$, where $u_\mathrm{0}$ is the impact parameter at the closest approach of the source and the lens. The method of constraining the angular Einstein radius, described in Section \ref{sec:method}, applies to such events. 

Additionally, given our primary interest in identifying dark lenses, we required $f_\mathrm{s}\geq0.8$ across most filters. Here, $f_\mathrm{s}$ is the blending parameter, which accounts for the potential influence of lens light on the light curve. Section \ref{sec:blending} provides a more detailed description of this parameter.

The astrometric parameters for the sources in the selected events are presented in Table \ref{tab:astrometry}. They come from Gaia Data Release 3 (GDR3) \citep{GDR3}.

\subsection{Gaia photometry} \label{sec:gaia-phot}
For all events considered, Gaia jointly collected 654 data points in the G filter \citep{2010JordiGaiaPhot}. Three of these events -- Gaia21efs, Gaia21azb, and Gaia21dpb -- were discovered through the Gaia Science Alerts system (GSA) \citep{Wyrzykowski2012, Hodgkin2013, Hodgkin2021}. GaiaDR3-ULENS-067 is from the first Gaia catalogue of candidate microlensing events, based on the GDR3 data \citep{Wyrzykowski2022, GDR3}.

For the events from GSA, the magnitude in the G filter varies from 19.800$^m$ (baseline magnitude for Gaia21dpb) to 12.180$^m$ (maximum magnitude for Gaia21efs). While magnitude errors are available in GDR3, they are not provided for GSA data. To estimate the errors, we used the formula described in \cite{Kruszynska2022}. Accordingly, the error bar for G = 19.800$^m$ is 0.077$^m$, and for G = 12.180$^m$ it is 0.003$^m$. The maximum and minimum magnitudes for GaiaDR3-ULENS-067 event from GDR3 are G = 17.053$^m\pm$0.007$^m$ and G = 14.557$^m\pm$0.002$^m$, respectively.

\subsection{Ground-based photometric observations} \label{sec:other-phot}

Gaia data alone are not sufficient to accurately constrain the FS effect in the microlensing events, as it observed each object roughly once a month \citep{Gaia}. To achieve a proper fit, the light curve must be covered as densely as possible. Fortunately, other photometric surveys and numerous ground-based observatories have collected data for the events considered in this study. We briefly describe them in this section.

The Zwicky Transient Facility (ZTF) is one of the major contributors to the data collection for our study \citep{ZTF}. This optical time-domain survey uses the 1.2 m telescope at Palomar Observatory, scanning the northern sky every two days. The survey collects data in the ZTF g, r, and i filters. For our study, ZTF provided 7079 points for the events Gaia21efs, Gaia21azb, and Gaia21dpb.

The Asteroid Terrestrial-impact Last Alert System (ATLAS) also contributed data to our study. This survey comprises four 50 cm telescopes located in Hawaii, Chile, and South Africa, allowing it to scan the sky several times per night. ATLAS utilises a variety of filters, including c (cyan), o (orange), g, r, i, H, and OIII \citep{ATLAS_Tonry2018, ATLAS-Smith2020}. The data from the survey were downloaded from the ATLAS Forced Photometry website \citep{Shingles2021}. For our analysis, ATLAS provided a total of 6168 data points in the o, c, i, g, r, and H filters for the events Gaia21efs, Gaia21azb, and GaiaDR3-ULENS-067.

Data from the Optical Gravitational Lensing Experiment (OGLE) are also available for GaiaDR3-ULENS-067. This survey is focused on dense regions such as the Galactic Bulge and the Magellanic Clouds. It uses a 1.3 m telescope located at Las Campanas Observatory in Chile \citep{Udalski2015}. Microlensing and other transient events are detected using OGLE Early Warning System (EWS) \citep{OGLE_EWS}. For GaiaDR3-ULENS-067, OGLE provided 3169 data points in OGLE I filter. 

Additional surveys that observed the sources include the Panoramic Survey Telescope and Rapid Response System 1 (Pan-STARRS1, PS1) and the DECam Plane Survey (DECAPS), providing high-quality photometry \citep{PS1,DECAPS}. PS1 data from 2011-2012 collected in the PS1 g, r, i, and z filters are available for Gaia21efs, Gaia21azb, and Gaia21dpb. Although these points were not used in modelling the microlensing curve, they were crucial for constraining the baseline magnitudes in filters where the baseline was not directly observed, and for estimating the angular radius of the sources, as described in Sections \ref{sec:modelling} and \ref{sec:source star}. DECAPS also provided precise photometry for GaiaDR3-ULENS-067 from 2017. These data were not used in our analysis, as a different method of source radius estimation was employed for this event (Section \ref{sec:GaiaDR3-ULENS-067}).

Data from infrared space telescopes, including the Two Micron All Sky Survey (2MASS) and the Near-Earth Object Wide-Field Infrared Survey Explorer (WISE/NEOWISE) \citep{2MASS, NEOWISE}, were also available. However, they were not utilised in the analysis due to their sparse coverage.

Lastly, the most numerous contribution came from the follow-up observations of the events from GSA -- 27 ground-based telescopes worldwide contributed a total of 8043 data points in UBVRI and ugriz filters. All of them are listed in Table \ref{tab:telescopes}. The telescope apertures vary from 30 cm to 2 m. To process and manage this data, the Black Hole TOM (BHTOM\footnote{\href{https://bhtom.space}{bhtom.space}}) was employed. Images from the ground-based telescopes corrected for bias, dark current, and flat field were uploaded to this platform. It facilitated the extraction of instrumental photometry for the target object as well as other objects in the field. The photometry was then automatically standardised to Gaia Synthetic Photometry (GaiaSP) \citep{GaiaSP} as is described in \cite{Zielinski2019, Zielinski2020}.

\subsection{Data cleaning} \label{sec:data cleaning}

Given the diverse data we have, data cleaning is an essential step to obtain accurate microlensing models. In this section, we give an overview of the data cleaning process for each event. All the photometric data available are listed in Tables \ref{tab:obs-Gaia21azb}-\ref{tab:obs-Gaia21efs}. Note that Gaia G data are available for every event and do not require pre-processing. Also, we do not mention data from infrared telescopes, or from PS1 and DECAPS, as they were not used in light curve modelling. Before any processing, we discarded possible duplicates. After cleaning, we scaled the data so that $\frac{\chi^2}{dof} = 1$ for each filter.

Gaia21efs. For this event, data from ZTF, ATLAS, and 20 ground-based telescopes are available. ZTF observations include data in all three filters: ZTF g, r, and i. We used only ZTF g and r data, as ZTF i data cover only the baseline. The majority of data points in the selected filters exhibit a characteristic curve on a magnitude-error plot, stemming from the Poissonian statistics of light. To ensure data uniformity, we removed points that were visually identifiable as obvious outliers in this plot. These outliers constitute <15\% of the r filter data points and <5\% of the g filter data points; thus, their removal does not significantly impact the curve coverage.

ATLAS data are available in ATLAS o, c, and H filters. We did not use the ATLAS H filter due to the small number of points. Some baseline data points in ATLAS o and c filters have significant error bars. Hence, we truncated the error distribution on the 75th percentile for each filter. Given the substantial datasets (>500 points for the c filter and >1800 points for the o filter), this still left us with a sufficient number of data points in the baseline.

The remaining data are available in filters U, B, V, R, I, u, g, r, i, and z. We excluded U, u, and z filters from our analysis because of the limited number of points. We also did not use data from TRT-SRO-0.7\_Andor-934. Despite the abundance of data points from this telescope, they have higher error bars compared to other observations at the same magnitudes. The remaining data, even after cleaning, adequately covers the curve. For the B filter, we truncated the error distribution at the 95th percentile to eliminate outliers. We applied the same procedure to the r filter. For the g and i filters, we exclusively utilised LCO data due to their smaller error bars and predominant representation in the dataset.

Data in V, R, and I filters come from a large set of data sources, so the cleaning technique we applied to them was rather unconventional. Firstly, we manually removed several apparent outliers. Then, for each filter, we divided the data roughly at the midpoint of the light curve. Depending on the extent of the distribution tails, we truncated them at either the 75th or 90th percentile. This approach ensures good-quality data near the peak, while maintaining sufficient data in the lower parts of the curve and the baseline.

Gaia21azb. Data for this event include observations from ZTF, ATLAS, and 12 ground-based observatories. We used ZTF g and r data without additional pre-processing and excluded ATLAS data due to high error bars at the given brightness. The remaining data are in the U, B, V, R, I, u, g, r, and i filters. We excluded the U and u filters due to the small number of points, and the B filter because of its large error bars and sparse coverage. The V filter was also excluded, as it appears to exhibit a systematic bias that adversely affected the model fit.

For the g, r, and i filters, we excluded the data points with the highest errors, truncating the error distribution of each filter at either 90th or 95th percentile, depending on its shape. The error bars of the V, R, and I filters are more significant than for the rest of the data. To make these datasets more uniform, we took the data only from the bigger telescopes (ASV1.4\_Andor, TJO\_MEIA2, TRT-SBO-0.7, pt5m). The error distributions of the selected data were additionally truncated at the 90th or 95th percentile, depending on their shapes.

GaiaDR3-ULENS-067. For this event, only data from surveys, namely from OGLE and ATLAS, are available. We used only OGLE data points without additional pre-processing, as they have small error bars, no significant outliers, and adequately cover both the light curve and the baseline.

Gaia21dpb. This event was observed by ZTF, LOIANO1.52\_BFOSC, HAO68\_G2-1600, and PIWNICE90\_C4-16000EC. We used ZTF g and r data without additional pre-processing. The remaining data mostly covered the baseline and could not be used to model the microlensing curve.

\subsection{Spectroscopic follow-up of Gaia21efs} \label{sec:spectroscopy}
In addition to photometry, Gaia21efs was bright enough -- especially during the peak magnification -- to obtain high-resolution spectroscopic data by using the Potsdam Echelle Polarimetric and Spectroscopic Instrument (PEPSI, \cite{Strassmeier2015}) mounted at the 2x8.4-m Large Binocular Telescope (LBT)\footnote{\href{https://www.lbto.org/}{www.lbto.org}} which is located on Mt. Graham, Arizona, US. The event was observed on November 3, 2021. The PEPSI configuration of a 300 $\mu$m fiber diameter and two cross-dispersers, CD2 (blue spectral arm) and CD5 (red spectral arm), were used simultaneously, which gave us a high signal-to-noise ratio (218) and a resolving power of R$\approx$40 000. After the standard calibration process (SDS4PEPSI, Ilyin 2000, PhD thesis, Univ. Oulu), we had obtained spectroscopic data divided into two parts: blue, covering the wavelength range between 422 and 478 nm, and red, covering the wavelength range between 624 and 743 nm.

\section{Methodology} \label{sec:method}

In this section, we describe the methodology we applied to measure the lower limit of the angular Einstein radius in every event studied. It comprises several steps. 

First, we used the MCMC technique to explore the parameter space of the combined model described in Section \ref{sec:phot-model} and derived posterior distributions for its parameters. The modelling process is outlined in Section \ref{sec:modelling}.

Next, we calculated the angular source radius, $\theta_{\ast}$, using colour-angular diameter relations, as is described in Section \ref{sec:source star}. Using this, we transformed a posterior distribution of $\rho$ (the angular size of the source relative to the angular Einstein radius) into the Einstein radius distribution, from which we estimated the lower limit of this parameter. This forms the basis for further analysis and interpretation. 

Then, we employed the \texttt{DarkLensCode} (DLC) software to estimate the lens mass, the lens distance, as well as the probability of the lens being a dark object (Section \ref{sec:dlc}). Finally, using the angular Einstein radius limit and the DLC results, we aimed to draw conclusions regarding the possible nature of the lensing object.

\subsection{Photometric model} \label{sec:phot-model}

We used a finite source point lens (FSPL) model in the vicinity of the peak and the point source point lens model (PSPL) in the remaining parts of the curve for all events. The specific time range for the FSPL model is given in Section \ref{sec:modelling}. Both models incorporate the annual microlens parallax. We used the PSPL model with a parallax as the initial model at the stage of event selection (Section \ref{sec:data}).

The expression for the photometric curve magnification, $A(u)$, in the case of the point source is as follows \citep{Paczynski1986}:

\begin{equation}
    A(u) = \frac{u^2+2}{u \sqrt{u^2+4}}
    \label{A(t) PSPL}
.\end{equation}

Here, $u$ is the angular separation between the lens and the source in the units of angular Einstein radius in a given moment. It is usually represented as the combination of components $\tau$ and $\beta$:

\begin{equation}
    u(t) = \sqrt{\tau(t)^2+\beta(t)^2} 
.\end{equation}

For the PSPL model with the annual parallax, the components are given by

\begin{equation}
    \tau(t) = \frac{(t-t_\mathrm{0})}{t_\mathrm{E}} + \delta \tau
    \label{tau_par}
\end{equation}
and
\begin{equation}
    \beta(t) = u_\mathrm{0} + \delta \beta
    \label{beta_par}
.\end{equation}

In these expressions, $u_\mathrm{0}$ represents the impact parameter, defined as the minimum separation between the source and the lens, occurring at the time of the light curve maximum, $t_\mathrm{0}$, and $t_\mathrm{E}$ denotes the timescale of the event. For each event, we find two very similar models corresponding to $u_\mathrm{0}>0$ and $u_\mathrm{0}<0$, as it is a common degeneracy. $\delta \tau$ and $\delta \beta$ are the corrections necessary to take into account the annual parallax. They are given by the following expression \citep{Gould2004}:

\begin{equation}
    (\delta \tau, \delta \beta) = (\vec{\pi}_\mathrm{E} \cdot \Delta \vec{s}, \vec{\pi}_\mathrm{E} \times \Delta \vec{s}), 
\end{equation}

where $\vec{\pi}_\mathrm{E}$ is the microlens parallax vector. Its northern and eastern components, denoted by $\pi_\mathrm{EN}$ and $\pi_\mathrm{EE}$, respectively, are included in the model. $\Delta \vec{s}$ denotes the positional offset of the Sun in the geocentric frame of reference.

We selected events with medium-to-long timescales, where parallax effects are typically detectable. To ensure that the parallax signal is not an artefact introduced by a single dataset, we compared the $\chi^2$ values for models with and without a parallax separately for each filter. In most cases, the inclusion of parallax led to a significant improvement in fit quality, with $\Delta \chi^2>5$ in the majority of filters, and $\Delta \chi^2>10$ in several -- indicating medium to strong evidence of the presence of a parallax across multiple datasets.

\subsubsection{Finite source effect} \label{FS_subsetion}

To incorporate the finite source (FS) effect, we integrated Formula \ref{A(t) PSPL} over the area of the source \citep{WittMao1994}:

\begin{equation}
    A_\mathrm{FS}(u) = \frac{1}{\pi \rho^2} \int_{S_\mathrm{source}}A(u)dS_\mathrm{source}
.\end{equation}

The parameter $\rho$ is included in the model and is given by

\begin{equation}
    \rho = \frac{\theta_{\ast}}{\theta_\mathrm{E}}
    \label{rho}
.\end{equation}

In this equation, $\theta_{\ast}$ represents the angular radius of the source. Typically, the FS effect becomes significant when the source size and the source-lens separation are comparable, i.e. when $\rho \approx u_\mathrm{0}$. In our events, however, we find that $\rho < u_\mathrm{0}$, meaning that $\rho$ cannot be directly measured but only constrained.

We employed the high-magnification approximation method outlined in \cite{Gould1994}. It simplifies the magnification for small $u$ to $A(u) \approx \frac{1}{u}$, facilitating easier integration. This approach is computationally efficient and remains sufficiently accurate for our analysis. Comparing its performance near the peak with the more precise FSPL method outlined in \cite{Lee2009}, we found that the most pronounced difference between the models is $\sim0.001\%$,  with a root mean square error of $\sim1.5 \cdot 10^{-3}$ in proximity to the peak.

We do not include limb darkening in our model, as we focus on events with $\rho < 0.1$. In these instances, it would not significantly alter the results, but it would introduce unnecessary complexity.

\subsubsection{Blending} \label{sec:blending}

Microlensing light curves can be affected by additional light that does not originate from the microlensed source. The most common causes include unresolved nearby sources in dense fields or the luminous lens. Therefore, the total flux, $F(t)$, is given by the expression

\begin{equation}
    F(t) = A(t) F_\mathrm{S}+ F_\mathrm{B} ,
    \label{flux}
\end{equation}

where $F_\mathrm{S}$ represents the source flux and $F_\mathrm{B}$ denotes the blend flux, originating from sources other than the microlensed one. We calculated them separately for every filter and for each set of model parameters using the \texttt{MulensModel} Python package described below.

To quantify the contribution of the source flux to the total flux, the blending parameter, $f_\mathrm{s}$, is typically introduced. It is defined as the fraction of the source flux in the total detected flux \citep{WozniakPaczynski1997}:

\begin{equation}
    f_\mathrm{s} = \frac{F_\mathrm{S}}{F_\mathrm{S}+F_\mathrm{B}} 
    \label{blending_parameter}
.\end{equation}

For events with negligible blend flux (i.e. with $f_\mathrm{s}\approx1$), we assume that all the light detected comes from the microlensed sources. Therefore, when the blending is close to unity in most filters for a particular event, we consider that the lens there is most likely dark.

With the source and blend fluxes, we calculated the apparent magnitude at baseline, denoted as $m_\mathrm{0}$, using the formula

\begin{equation}
    m_\mathrm{0} = 22.0^m - 2.5log(F_\mathrm{S}+F_\mathrm{B}) ,
    \label{apparent_magnitude}
\end{equation}

where $22.0^m$ is assumed to be a zero point value.

\subsection{Modelling} \label{sec:modelling}

In summary, the model we used includes the following parameters: ($t_\mathrm{0}$, $u_\mathrm{0}$, $t_\mathrm{E}$, $\pi_\mathrm{EN}$, $\pi_\mathrm{EE}$, $\rho$) and pairs ($F_\mathrm{S}$, $F_\mathrm{B}$) for each filter. We applied the MCMC method using the \texttt{emcee} Python package, which utilises an ensemble sampler, allowing us to explore parameter space in parallel with multiple chains (‘walkers’) and achieve faster convergence and better sampling accuracy \citep{EMCEE}. 

We also used the \texttt{MulensModel} package to streamline the handling and analysis of microlensing data \citep{MulensModel}. It is helpful in particular in fitting ($F_\mathrm{S}$, $F_\mathrm{B}$) pairs for every model sampled. Additionally, in \texttt{MulensModel}, various methods for calculating magnification are available for both point and finite source scenarios. We employed the \texttt{finite\_source\_uniform\_Gould94} method in the vicinity of the peak, specifically during the time range corresponding to $u \leq 3u_\mathrm{0}$ \citep{Gould1994}. This is the implementation of the Gould approximation described in Section \ref{FS_subsetion}. For the remainder of the curve, we employed the \texttt{point\_lens} method, which is the default option.

In Gaia21efs and Gaia21azb, many filters lack sufficient baseline coverage, which leads to blending values exceeding one -- unphysical in the standard microlensing formalism. To mitigate this, we applied Gaussian priors on the baseline magnitudes for selected filters: 

\begin{equation}
    \mathrm{ln(Prior)} = -\frac{1}{2} \left( \frac{m_0 - m_{\mathrm{exp}}}{\sigma} \right)^2
.\end{equation}

Here, $m_0$ is the baseline magnitude inferred from the model, $m_{\mathrm{exp}}$ is the expected baseline magnitude based on independent photometry, and $\sigma$ represents the uncertainty set to 1–5\% of the expected magnitude. The expected baseline magnitudes for the BVRI and gri bands were derived from PS1 photometry, using the colour transformations provided by \citet{Tonry2012}. Additionally, we applied a Gaussian prior to the blend flux:

\begin{equation}
  \mathrm{ln(Prior)} = -\frac{1}{2} \left( \frac{F_{\mathrm{B}}}{0.05} \right)^2
.\end{equation}

This prior favours low blend flux values without imposing a hard constraint of zero. More details on using these priors can be found in Sections \ref{sec:Gaia21efs} and \ref{sec:Gaia21azb}.

The filters lacking baseline coverage include observations from both large and small telescopes. Therefore, anomalous blending values in these bands could also be a result of unresolved nearby sources or photometric systematics in the lower-resolution data from smaller telescopes. We tested models excluding data from small telescopes, but blending did not improve noticeably without them. Therefore, we retained these datasets to preserve good coverage of the curve.

The posterior distributions of all parameters except $\rho$ typically exhibit Gaussian behaviour in our analysis. We characterised these distributions by determining the median as the parameter value. Subsequently, their uncertainties were quantified as the differences between the medians and the 16th or 84th percentiles. However, in cases in which the distribution displays noticeable skewness, we employed an alternative procedure. Specifically, the parameter value is a mode of distribution, and its uncertainties are the differences between the mode and the boundaries of the highest density interval, encompassing 68\% of the samples.

The posterior distribution of $\rho$, in turn, tends to be concentrated near 0. We characterised this distribution by the upper limit, further denoted as $\rho_\mathrm{lim}$. We defined it to be at the 95th percentile of its posterior distribution, corresponding to a one-sided Bayesian credible interval. The 95\% threshold is a common and statistically sound choice, excluding the least probable tail of the distribution without being overly conservative. It is worth noting that this limit may be sensitive to the starting value of $\rho$ in the MCMC process. To prevent it from being underestimated, we set the initial value of $\rho$ to be approximately equal to $u_0$. This is the highest reasonable value, as if the FS effect is not present, the $\rho$ distribution is expected to be smaller than $u_0$.

\subsection{Source stars} \label{sec:source star}

We used parallaxes and proper motions of the sources published in GDR3. These parameters are listed in Table \ref{tab:astrometry}. Our analysis also requires their angular radius, $\theta_{\ast}$, distance, $D_\mathrm{S}$, and extinction in the G filter, $A_\mathrm{G}$. We now describe how we obtained these parameters.

\subsubsection{Angular radius}

We computed the angular radius using empirical colour-angular diameter relations provided in \cite{Adams2018}. These relations are constructed for giants or dwarfs and subgiants. To discern the type of microlensed source, we constructed colour-magnitude diagrams (CMDs) using objects from the GDR3 within $5'$ circles around the source. Based on them, we treated the sources in Gaia21efs, Gaia21azb, and GaiaDR3-ULENS-067 as giants and the source in Gaia21dpb as a main-sequence star. For Gaia21efs, this classification is consistent with the spectroscopic analysis. Then, we employed the appropriate relations to calculate the source sizes. 

We calculated colours for Gaia21dpb, Gaia21azb, and Gaia21efs using PS1 photometry. We corrected these for extinction using values for PS1 filters available in \cite{Schlafly2011}. Next, we converted PS1 filters to the UBVRI system. We did this by applying relations from \cite{Tonry2012}. For the source of GaiaDR3-ULENS-067, a different set of filters and procedures were employed, as it is a red clump (RC) star from the Bulge. This is detailed in Section \ref{sec:GaiaDR3-ULENS-067}.

Given the posterior distribution of $\rho$ and an estimate of the source angular radius, $\theta_*$, we derived a lower limit on the Einstein radius, $\theta_\mathrm{E,lim}$. To avoid unphysically large $\thetaE$ values caused by the few extremely small $\rho$ samples, we discarded the lowest $\rho$ values. The cut-off was placed where the number of samples begins to rise rapidly, indicating the start of the well-constrained portion of the posterior. We then generated a Gaussian distribution of $\theta_*$ with the same number of samples as $\rho$, accounting for its measurement uncertainty. The resulting $\thetaE$ distribution was computed via sample-wise division (transformation of Formula \ref{rho}). $\theta_\mathrm{E,lim}$ was then taken as the fifth percentile of this distribution, corresponding to a one-sided 95\% lower credible interval, analogous to $\rho_{lim}$ estimation.

\subsubsection{Distance}

The distances to the sources were determined based on various considerations specific to each event. When the RUWE of the Gaia astrometric solution approximately equals unity, we can reasonably rely on Gaia parallaxes. This applies to the sources of Gaia21efs and Gaia21azb (RUWE values are given in Table \ref{tab:astrometry}). Accordingly, for Gaia21azb, we adopted the distance value from \cite{Bailer-Jones2018, 2021Bailer-Jones}. In contrast, for Gaia21efs, we used a spectroscopic distance derived from a high-resolution spectrum. For GaiaDR3-ULENS-067 and Gaia21dpb, we used other approaches, described in Sections \ref{sec:GaiaDR3-ULENS-067} and \ref{sec:Gaia21dpb}, respectively. Below, we describe how the atmospheric parameters and spectroscopic distance for Gaia21efs were obtained.

We analysed a high-resolution spectrum from LBT/PEPSI using the iSpec\footnote{\href{https://www.blancocuaresma.com/s/iSpec}{www.blancocuaresma.com/s/iSpec}} framework \citep{BlancoCuaresma2014,BlancoCuaresma2019}. The analysis employed the SPECTRUM\footnote{\href{http://www.appstate.edu/~grayro/spectrum/spectrum.html}{www.appstate.edu/~grayro/spectrum/spectrum.html}} radiative transfer code, MARCS model atmospheres \citep{Gustafsson2008}, and solar abundances from \citet{Grevesse2007}.

\begin{figure}
    \centering
    \includegraphics[width=\linewidth]{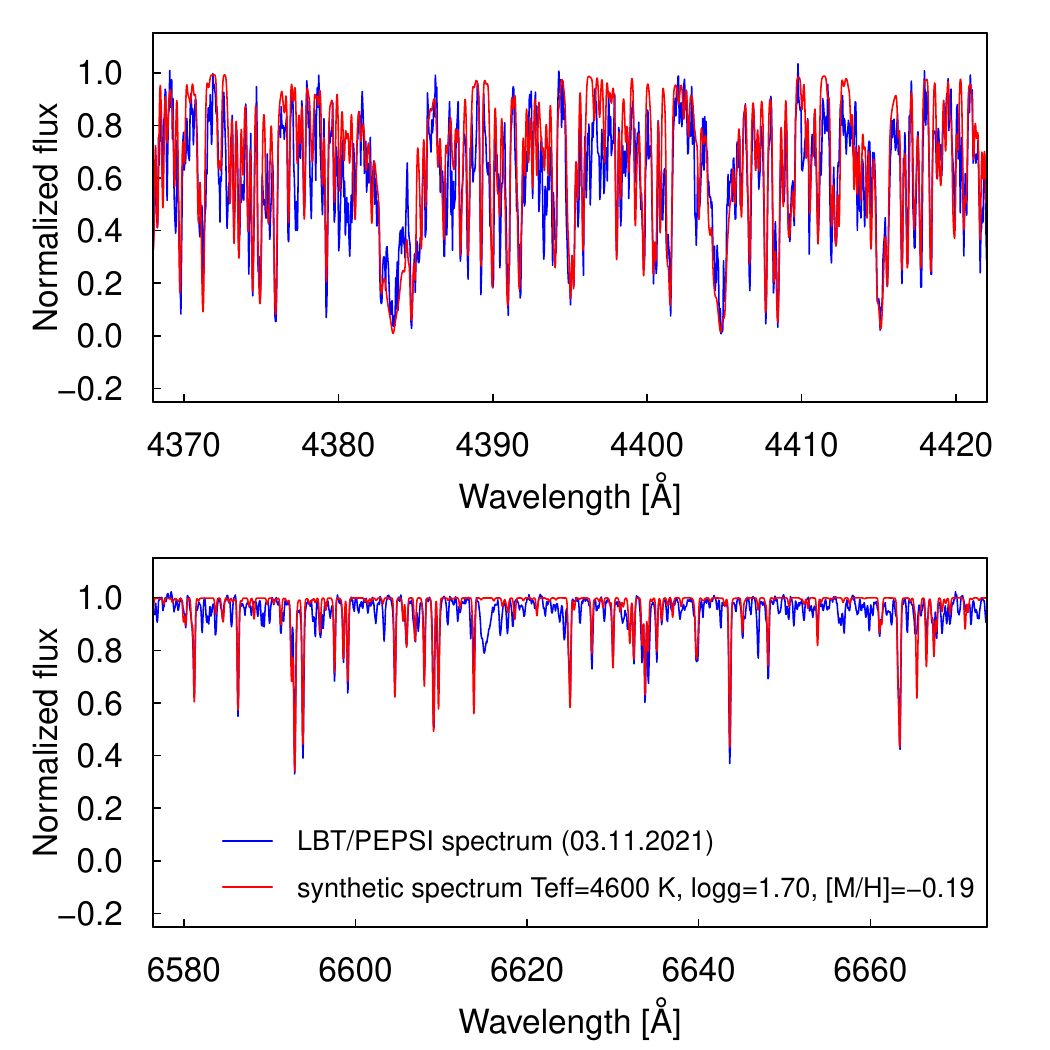}
    \caption{Spectrum of the Gaia21efs obtained with LBT/PEPSI on November 3, 2021 (blue) and the
best-matching fit (red) synthesised for the specific parameters. The Fe I regions in blue (top) and red spectral arms (bottom) are presented.}
    \label{fig:spectrum}
\end{figure}

Figure \ref{fig:spectrum} presents the observed LBT/PEPSI spectrum together with the best-matching spectrum synthesised for the following atmospheric parameters: effective temperature $T_{\rm eff}$ = 4600 $\pm$ 22 K, surface gravity $\log g$ = 1.70 $\pm$ 0.07, metallicity [M/H] = -0.19 $\pm$ 0.03 dex, and microturbulence velocity $v_{\rm t}$ = 1.26 $\pm$ 0.02 km s$^{-1}$. Based on this solution, we classify the source star in Gaia21efs event as G8 II spectral type and estimate its absolute magnitude to be M$_{\rm V}$=0.9$^m$ $\pm$ 0.3 \citep{Straizys1992}. Assuming an apparent magnitude of V=16.546$^m$ (from the microlensing model) as well as a line-of-sight extinction of A$_{\rm V}$=2.235$^m$ \citep{Schlafly2011}, we were able to estimate the spectroscopic distance to the source star as D$_{S}$=4.81 $\pm$ 0.91 kpc. The distances presented in \cite{Bailer-Jones2018, 2021Bailer-Jones} are in agreement with this value within 1$\sigma$ uncertainty. Therefore, in the case of Gaia21efs, the above spectroscopic distance was adopted in order to constrain the distance and mass to the lens in this specific microlensing event.

\subsubsection{Extinction}

Extinction in the G filter is provided in the GDR3 as the \texttt{ag\_gspphot} parameter. It is available only for one of the sources considered in this work:  Gaia21azb. For Gaia21efs and Gaia21dpb, we adopted a similar procedure as is described in \cite{Kruszynska2022}. Specifically, we calculated the average \texttt{ag\_gspphot} extinction from approximately four nearest sources within $20''$ circle around the source. Finally, we computed extinction for GaiaDR3-ULENS-067 using the CMD (Section \ref{sec:GaiaDR3-ULENS-067}).

\subsection{\texttt{DarkLensCode}} \label{sec:dlc}

The \texttt{DarkLensCode} (DLC) is a software used to obtain the posterior distribution of the lens mass and distance, as well as to estimate its probability of being dark. It utilises the posterior distributions of the photometric model parameters and a Galactic model. DLC is based on the approach described by \cite{Wyrz16}, \cite{MrozWyrzykowski2021} and is detailed in \cite{Howil2024}. Below, we briefly summarise the main steps of the DLC procedure.

We provide the event co-ordinates, the adopted mass function, and the posterior distributions of the parameters: $t_0$, $u_0$, $t_\mathrm{E}$, $\pi_\mathrm{EN}$, $\pi_\mathrm{EE}$, $m_{0,G}$, and $f_{S,G}$ (where the subscript G denotes quantities related to the Gaia G filter). Additionally, the source distance, $D_S$, components of the source proper motion, $\mu_S$, and extinction, $A_G$, are included as inputs.

The procedure begins by drawing a random sample from the posterior distribution. The relative proper motion of the lens and source, $\mu_{rel}$, is sampled from a uniform distribution in the range (0, 30) mas yr$^{-1}$, while $D_S$ is drawn from either a uniform or Gaussian distribution, depending on the configuration parameter \texttt{ds\_weight}. Using these values, the angular Einstein radius, $\theta_E$, lens mass, $M_L$, and lens distance, $D_L$, are computed for each sample. This process is repeated for the desired number of iterations. Each sample is subsequently weighted based on the Galactic priors. 

These priors include the relative proper motion, the lens distance, and the lens mass. The distance prior comes from \cite{HanGould2003} and \cite{Batista2011}. The prior for relative proper motion is assumed to be Gaussian, derived from separate Gaussian distributions for the source and lens proper motions. The source proper motion is characterised by an input mean value, $\mu_S$, and variance, $\sigma_S$ (see Table \ref{tab:astrometry}). The lens proper motion mean, $\mu_L$, is computed using standard kinematic relations from \cite{Reid2009}. The variance of the lens proper motion is determined by its position in the Galaxy: $(\sigma_l, \sigma_b) = (100, 100)$ km s$^{-1}$ for Bulge lenses and $(\sigma_l, \sigma_b) = (30, 20)$ km s$^{-1}$ for disc lenses \citep{Howil2024}.

The lens mass prior, i.e. the mass function, is another input that we define. Since the true mass function for dark stellar remnants remains uncertain, we explored several plausible assumptions. Specifically, we tested the standard Kroupa mass function for stars \citep{Kroupa2001}, an approximate mass function for stellar remnants from \cite{Mroz2021MF}, and a simple power-law form, $f(M) \propto M^{-1}$. Among these, the Kroupa mass function yields the lowest inferred lens masses and the most conservative estimates of dark lens probabilities. We therefore adopted it as our default in the analysis. However, if the actual mass function more closely resembles that proposed by \cite{Mroz2021MF}, the evidence supporting dark lens candidates would be even stronger.

The dark lens probability was computed by comparing the inferred brightness of the lens with the expected brightness of a main-sequence star of the same mass. The actual lens brightness distribution was derived from the blending and baseline magnitude posterior distributions provided as input. For each sample, the expected main-sequence brightness was calculated based on its estimated mass and distance, using empirical relations from \cite{PecautMamajek2013}. DLC computes the probability for zero extinction and for the extinction provided as an input, which is considered the maximum possible extinction. Consequently, we obtained a range for the dark lens probability for zero and maximum extinction.

As we have determined the lower limit for the angular Einstein radius, we included only the samples for which $\theta_\mathrm{E}\geq\theta_\mathrm{E,lim}$ in the resulting posterior distributions of mass, distance, and lens brightness. From these filtered samples, we computed the median mass and distance, as well as the dark lens probability.

It is also useful to calculate the transverse velocity of the lens $v_\mathrm{\perp,L}$, which is given by

\begin{equation}
    v_\mathrm{\perp,L} = \mu_\mathrm{L} D_\mathrm{L},
\end{equation}

where the proper motion of the lens $\mu_\mathrm{L}$ can be expressed as the vector sum $\vec{\mu}_\mathrm{L} = \vec{\mu}_\mathrm{rel} + \vec{\mu}_\mathrm{S}$ \cite{Gould2000b}. DLC provides us with the posterior distributions of $\mu_\mathrm{rel}$, $\mu_\mathrm{S}$, and $D_\mathrm{L}$. Since the microlensing parallax vector $\vec{\pi}_{\rm E}$ and the relative proper motion $\vec{\mu}_{\rm rel}$ are aligned, we can determine the angle between $\vec{\mu}_S$ and $\vec{\mu}_{\rm rel}$ using the dot product, allowing us to compute the full vector,  $\vec{\mu}_\mathrm{L}$\citep{GouldParallaxVector}. Combining all components, we derived the posterior distribution of the lens transverse velocity, which is weighted as the standard DLC posteriors. As before, we used the $\theta_{\rm E,lim}$ constraint for this distribution and computed its median.

\section{Results} \label{sec:results}

\begin{table*}[!htbp]
\small
\centering
\caption{Parameters of the events derived in this work.}
\label{tab:all results}
\begin{tabular}{c|cc|cc|cc|cc}
\hline
&	\multicolumn{2}{c|}{Gaia21efs}	&	\multicolumn{2}{c|}{Gaia21azb}			&	\multicolumn{2}{c|}{GaiaDR3-ULENS-067}		&	\multicolumn{2}{c}{Gaia21dpb}			\\ \hline								
Parameter	&	$u_0>0$	&	$u_0<0$	&	$u_0>0$	&	$u_0<0$	&	$u_0>0$	&	$u_0<0$	&	$u_0>0$	&	$u_0<0$	\\								
\hline
\hline
$\theta_{\ast}$ [$\mu as$]	&	\multicolumn{2}{c|}{$9.0\pm1.1$}			&	\multicolumn{2}{c|}{$3.2\pm0.4$}			&	\multicolumn{2}{c|}{$7.7\pm1.4$}			&	\multicolumn{2}{c}{$1.01\pm0.03$}			\\	

$D_\mathrm{S}$ [kpc]	&	\multicolumn{2}{c|}{{$4.81\pm0.91$}}		&	

\multicolumn{2}{c|}{$5.1^{+2.5}_{-1.4}$}			&	\multicolumn{2}{c|}{$8.0\pm2.0$}			&		\multicolumn{2}{c}{$8.61^{+2.23}_{-1.77}$}	\\					
$A_\mathrm{G}$ [mag] & \multicolumn{2}{c|}{1.66$\pm 0.09$} & \multicolumn{2}{c|}{0.95$^{+0.07}_{-0.08}$} & \multicolumn{2}{c|}{$1.74\pm0.24$} & \multicolumn{2}{c}{0.63$\pm 0.06$} \\

$\rho_\mathrm{lim}$ &	
0.0092 &			
0.0092	&	
0.0058	&	
0.0057	&	
0.0258	&	
0.0236	&	
0.0249	&	
0.0253	\\	

$\theta_\mathrm{E,lim}$ [$\mu as$]	&	
944.7	&	
947.1	&	

$538.1$	&	
$543.4$	&	

$271.6$	&	
$294.2$	&	

$40.4$	&	
$39.8$	\\	

$M_\mathrm{L, lim}$ [$M_{\odot}$]	&
$0.75^{0.08}_{-0.08}$	&	
$0.77^{+0.09}_{0.08}$	&	

$0.45^{+0.17}_{-0.14}$	&	
$0.51^{+0.18}_{-0.16}$	&

$0.42^{+0.08}_{-0.07}$	&	
$0.36^{+0.05}_{-0.05}$	&	

$0.0139^{+0.0003}_{-0.0003}$	&	
$0.0141^{+0.0003}_{-0.0003}$	\\		

$D_\mathrm{L,lim}$ [kpc]	&	
$3.2^{+0.3}_{-0.3}$	&	
$3.2^{+0.3}_{-0.3}$	&	

$3.6^{+1.3}_{-0.8}$	&	
$3.7^{+1.4}_{-0.8}$	&	

$6.8^{+1.5}_{-1.5}$	&	
$6.5^{+1.3}_{-1.3}$	&	

$7.66^{+1.77}_{-1.40}$	&	
$7.70^{+1.78}_{-1.4}$	\\			

$M_\mathrm{L, DLC}$ [$M_{\odot}$]	&
$0.97^{+0.45}_{-0.19}$	&	
$1.00^{+0.45}_{-0.20}$ 	&	

$1.04^{+0.73}_{-0.41}$	&	
$1.14^{+0.78}_{-0.44}$	&	

$1.70^{+0.83}_{-0.60}$	&	
$1.18^{+0.55}_{-0.39}$	&	

$0.43^{+0.39}_{-0.20}$	&	
$0.45^{+0.41}_{-0.21}$	\\

$D_\mathrm{L, DLC}$ [kpc]	&	
$1.9^{+0.7}_{-0.7}$	&	
$1.9^{+0.7}_{-0.7}$ &	

$3.5^{+0.9}_{-0.9}$	&	
$3.6^{+0.9}_{-0.8}$	&	

$7.2^{+1.3}_{-1.0}$	&	
$7.1^{+1.3}_{-1.0}$	&	

$1.8^{+1.1}_{-0.8}$	&	
$1.8^{+1.1}_{-0.8}$	\\								

DLP [\%]	&	
87-100	&	
89-100	&

86-94 &	
90-95	&	

80-94	&	
80-94	&	

82-87	&	
83-88	\\

$v_\mathrm{\perp,L}$ [km s$^{-1}$] &
$120^{+33}_{-31}$ &
$121^{+34}_{-32}$ &

$187^{+36}_{-35}$ &
$193^{+36}_{-35}$ &

$321^{+40}_{-40}$ &
$298^{+36}_{-34}$ &

$96^{+25}_{-19}$ &
$95^{+25}_{-17}$ \\

\hline

\end{tabular}

\tablefoot{$\theta_{\ast}$ -- angular radius of the source, $D_\mathrm{S}$ -- distance to the source, $A_\mathrm{G}$ -- extinction in the G filter, $\rho_\mathrm{lim}$ -- upper limit for source radius in the units of the Einstein radius, $\theta_\mathrm{E,lim}$ -- lower limit for the angular Einstein radius, $M_\mathrm{L,lim}$ -- lower limit for the lens mass as computed with $\theta_\mathrm{E}=\theta_\mathrm{E,lim}$, $D_\mathrm{L,lim}$ -- upper limit for the lens distance (analogous to $M_\mathrm{L,lim}$), $M_\mathrm{L,DLC}$ -- median lens mass from the DLC procedure if $\theta_\mathrm{E,lim}$ is taken into account, $D_\mathrm{L,DLC}$ -- median lens distance from the DLC procedure if $\theta_\mathrm{E,lim}$ is taken into account, DLP -- dark lens probability from the DLC, $v_\mathrm{\perp,L}$ -- transverse velocity of a lens.}
\end{table*}

In this section, we present the results of our analysis for each microlensing event. The key parameters, including the Einstein radius limit, lens mass, and distance, along with the dark lens probability from DLC, are summarised in Table \ref{tab:all results}. Additionally, we list the lower mass limit and upper distance limit of the lens, calculated using $\theta_\mathrm{E,lim}$ with Formula \ref{eq:mass} and Formula \ref{eq:distance}, respectively. 

We present light curves of the events and two types of plots generated by the DLC. The first is the mass-distance plot. The second is the blend-lens plot representing the blend light compared to the lens light, assuming it is a main-sequence star. The dashed line in the blend-lens plot indicates the extinction's lower limit, and the solid line shows its upper limit. The colours indicate the probability density on a logarithmic scale (base 10).

The bright samples in the DLC plots meet the condition $\theta_\mathrm{E} \geq \theta_\mathrm{E,lim}$. They are overlaid on the shaded samples, which represent the original distribution without considering this limit. All plots representing the light curve or the DLC results show only the model with $u_0 > 0$, as the models with $u_0 < 0$ yielded very
similar plots. In the figures in which the light curves are presented, both models FSPL and PSPL include the annual parallax.

\subsection{Gaia21efs} \label{sec:Gaia21efs}

\begin{figure}[h]
    \centering
    \includegraphics[width=8.5cm]{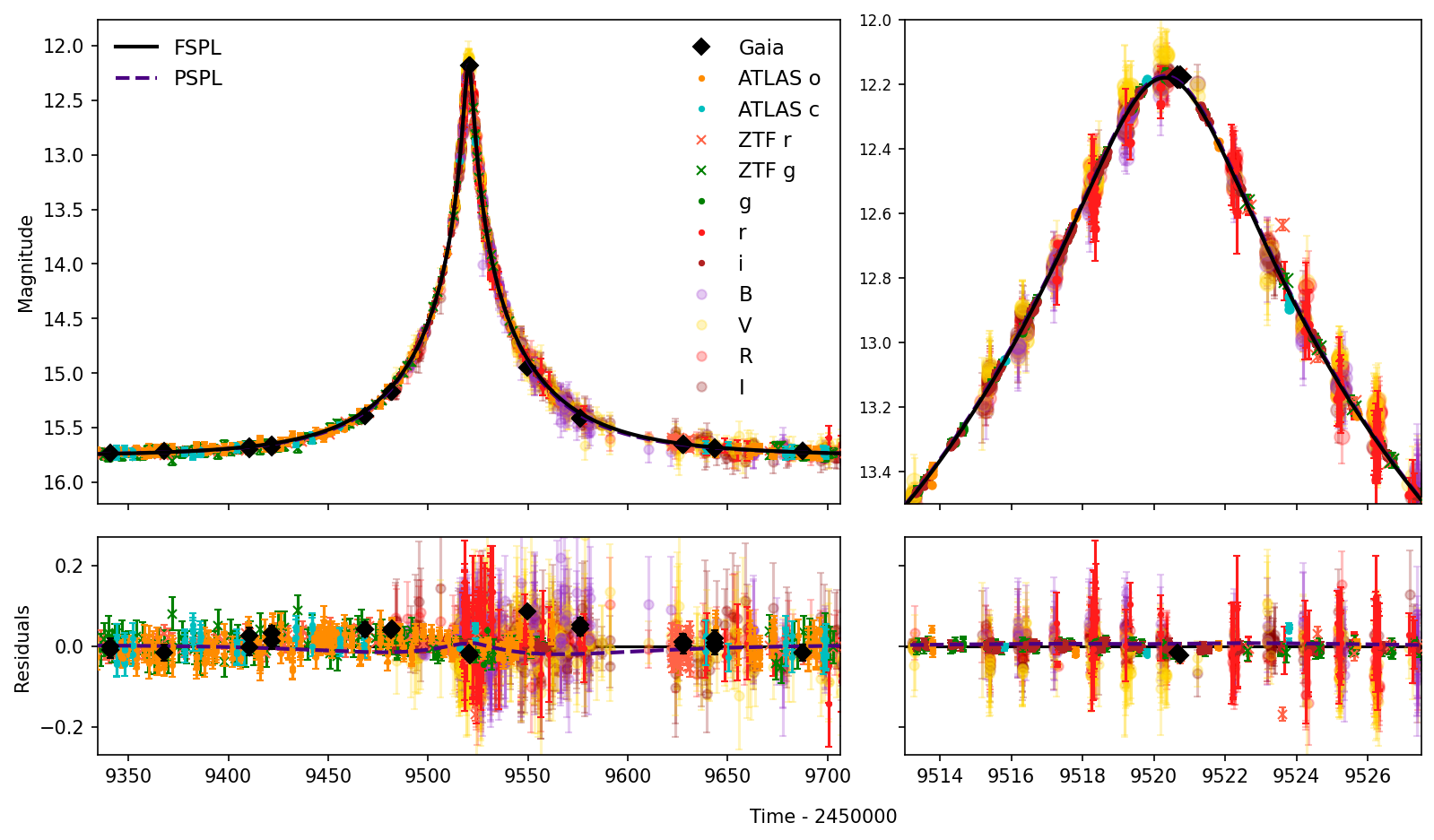}
    \caption{Light curve of Gaia21efs with the data from different surveys as well as the follow-up data collected with BHTOM. The solid line shows the parallax FS model, while the dashed line shows a parallax model for the point source. The right figure shows the zoom-in on the peak of the light curve.}
    \label{fig:Gaia21efs_lc}
\end{figure}

\begin{figure}[h]
\centering
\begin{subfigure}
  \centering
  \includegraphics[width=9cm]{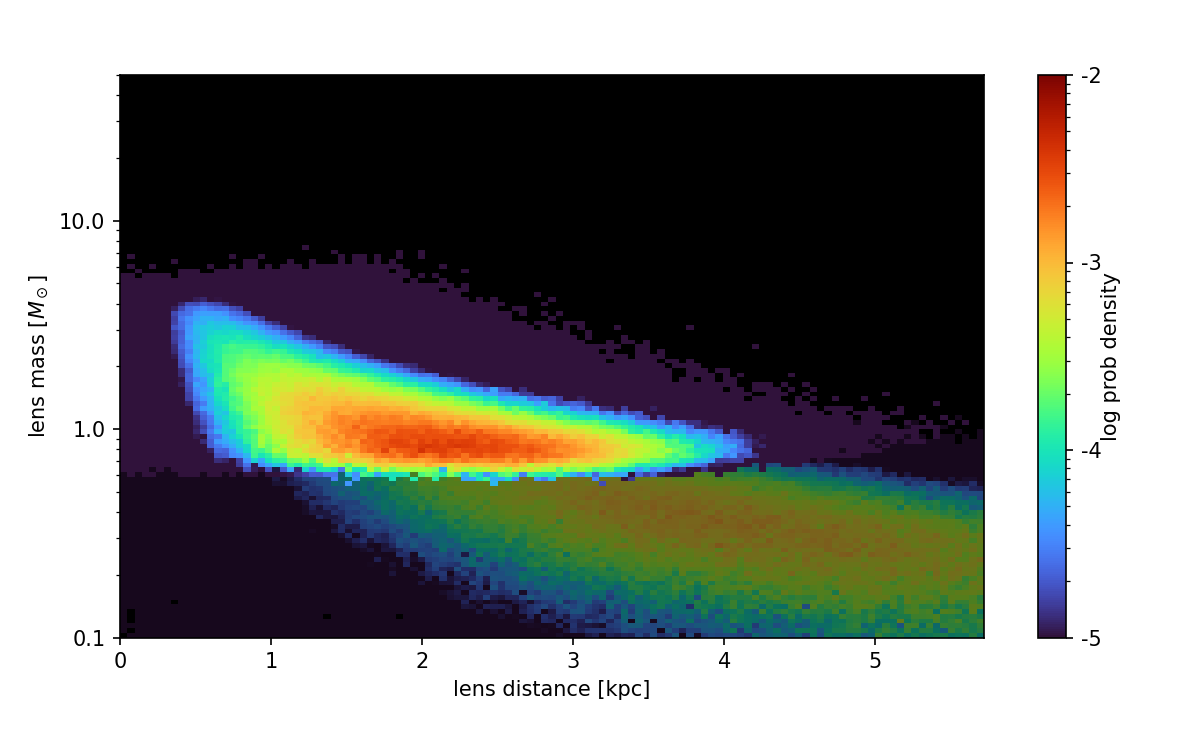}
\end{subfigure}

\begin{subfigure}
  \centering
  \includegraphics[width=8.33cm]{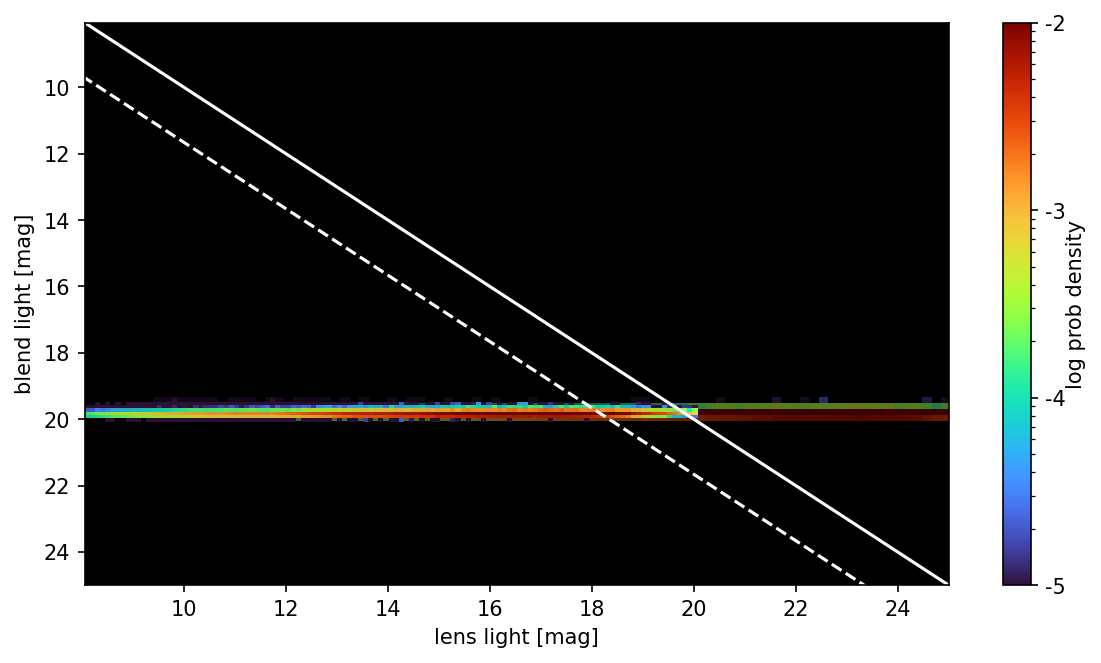}
\end{subfigure}
\caption{DLC results for Gaia21efs. Top: Mass-distance plot. Bottom: Blend-lens plot.}
\label{fig:Gaia21efs DLC}
\end{figure}

Initial modelling using all available datasets for this event resulted in unphysical blending values ($f_S>1$) in many filters, including Gaia G, likely due to the lack of baseline coverage in some of them. To investigate this, we initially modelled the event using only the datasets that provided good coverage of both the magnification peak and the baseline. Specifically, we began with the following filters: Gaia G, ZTF g, ZTF r, ATLAS o, and ATLAS c. This combination yielded a model with blending parameters close to unity and consistent with physical expectations. We refer to this as the base model, which serves as a reference point in our analysis; the inclusion of additional filters or priors is expected to preserve the core parameters of this model.

We then added the g- and i-band data from LCO, which, despite lacking a baseline, are homogeneous in quality and sample the peak well. With this subset (Gaia G, ZTF g, ZTFr, ATLAS o, ATLAS c, g, and i), physically plausible blending ratios were recovered by applying Gaussian priors on the blend flux and baseline magnitude in the ATLAS o, ATLAS c, g, and i bands, as is described in Section \ref{sec:modelling}. Although ATLAS data provide baseline coverage, their blending estimates remained sensitive to the addition of filters without baseline, and thus also required priors. The baseline magnitude prior means for the ATLAS filters were set to the values from the base model.

Finally, we included the remaining filters while maintaining the same priors and fixing blend flux in the r filter to zero. This allowed us to incorporate all available data without introducing unphysical blending. This constitutes the final model presented in this work. It differs from the base model at $\sim2\sigma$ level, particularly in the microlens parallax, which is sensitive to the additional filters.

The light curve of Gaia21efs is shown in Figure \ref{fig:Gaia21efs_lc}. The lens mass limit for this event exceeds 0.7 $M_{\odot}$, and the dark lens probability is over 85\%. The top panel of Figure \ref{fig:Gaia21efs DLC}, which displays the mass-distance plot, highlights the significant influence of this limit on the DLC results. Without taking it into account, the median mass from the DLC analysis would be around $\sim0.4$ $M_{\odot}$. In contrast, when considering this limit, the inferred mass is $\sim1.0$ $M_{\odot}$. This discrepancy suggests that such high-mass objects are relatively rare in this region of the Galaxy. This observation motivated us to estimate the transverse velocity of the lens, as described in Section \ref{sec:dlc}, as it could have travelled to this region from a different place in a Galaxy. A high velocity, potentially caused by a supernova kick, would provide additional evidence supporting its compact nature.

When taking into account the Einstein radius limit, the median transverse velocity of the lens is $\sim120$ km s$^{-1}$, with the lower bound of $\sim 90$ km s$^{-1}$. These values are comparable with the mean transverse velocity of recycled pulsars ($87\pm13$ km s$^{-1}$) reported by \cite{Hobbs:2005}, but significantly lower than of non-recycled pulsars ($246\pm22$ km s$^{-1}$). At the same time, \cite{Wegg2012} suggested that white dwarfs with masses greater than 0.75 $\msun$, as in our case, typically exhibit transverse velocities up to 60 km s$^{-1}$. Given that the lens velocity exceeds this range, it is consistent with either a relatively high-velocity white dwarf or a low-velocity neutron star.

\subsection{Gaia21azb} \label{sec:Gaia21azb}

Similarly to Gaia21efs, the initial model of Gaia21azb exhibited unphysical blending in some filters because of incomplete baseline coverage, so we applied a similar approach as for the previous event. In this case, the filters covering both the baseline and the curve are Gaia G, ZTF g, and ZTF r. The blending parameters for this combination were physically reasonable and close to one, forming the base model for this event.

\begin{figure}[!htbp]
    \centering
    \includegraphics[width=9cm]{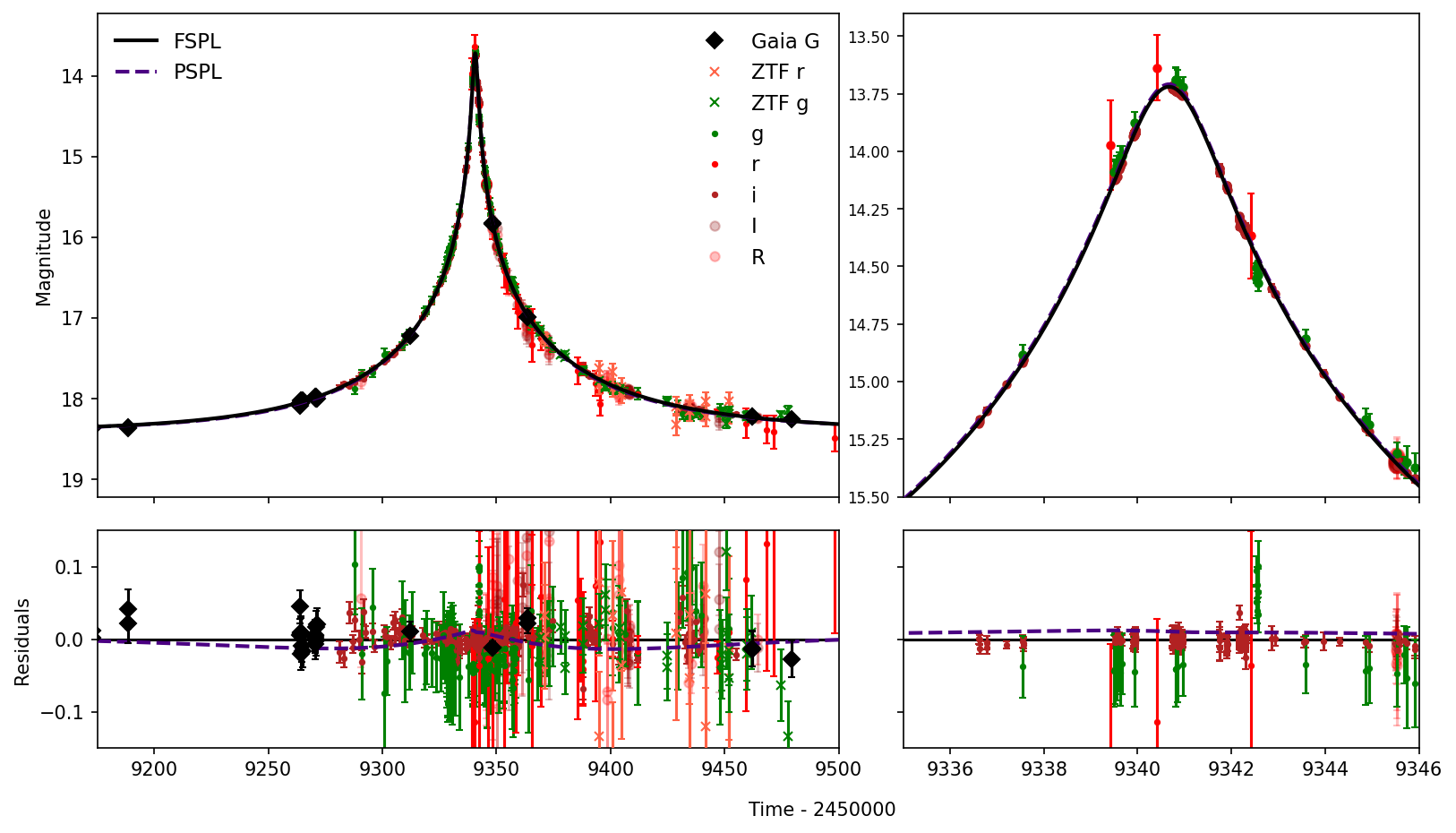}
    \caption{Light curve of Gaia21azb with the data from different surveys as well as the follow-up data collected with BHTOM. The solid line shows the parallax FS model, while the dashed line shows a parallax model for the point source. The right figure shows the zoom-in on the peak of the light curve.}
    \label{fig:Gaia21azb_lc}
\end{figure}

Subsequently, we modelled the event using an extended set of filters: Gaia G, ZTF r, and ZTF g, g, and i. The photometry in the g and i bands comes from LCO and was treated with priors as is described in Section \ref{sec:modelling}, leading to physically consistent blending values across all filters. Adding the remaining filters with the same treatment produced similar results, so we adopted this as our final model. Its parameters are consistent with the base model within 1$\sigma$, except for $u_0$ and $t_\mathrm{E}$, which agree within 3$\sigma$.

\begin{figure}[h]
\centering
\begin{subfigure}
  \centering
  \includegraphics[width=9cm]{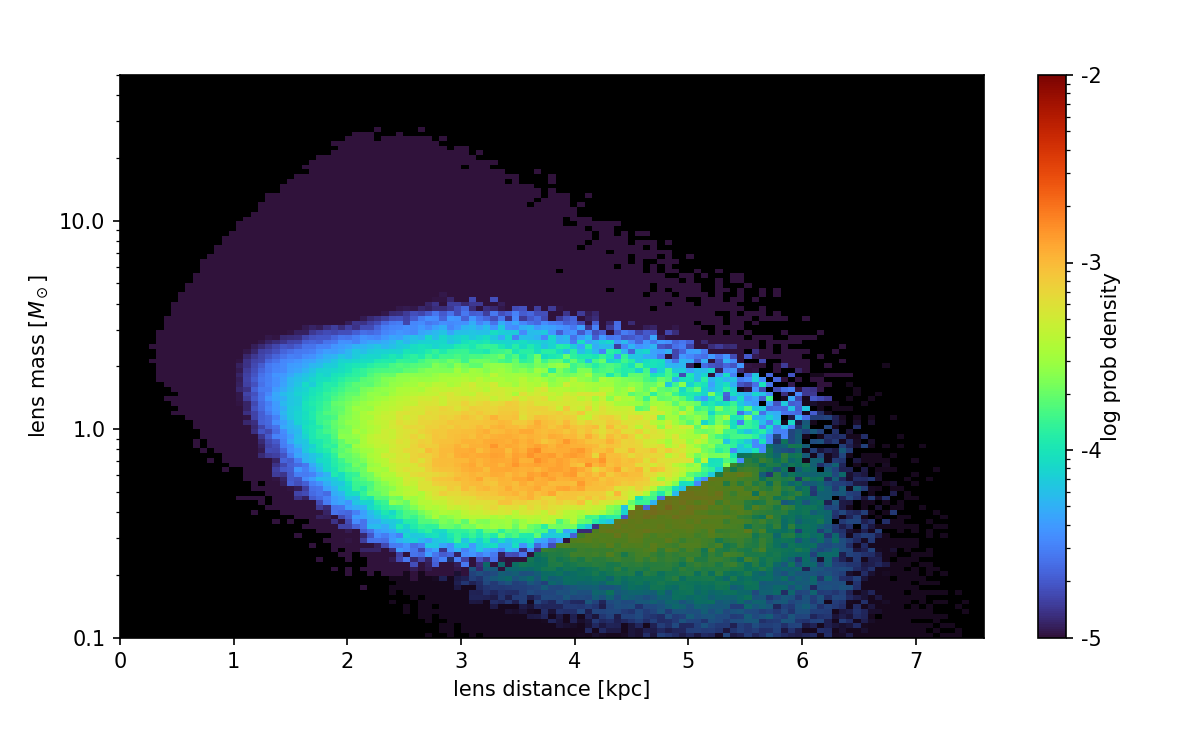}
\end{subfigure}

\begin{subfigure}
  \centering
  \includegraphics[width=8.33cm]{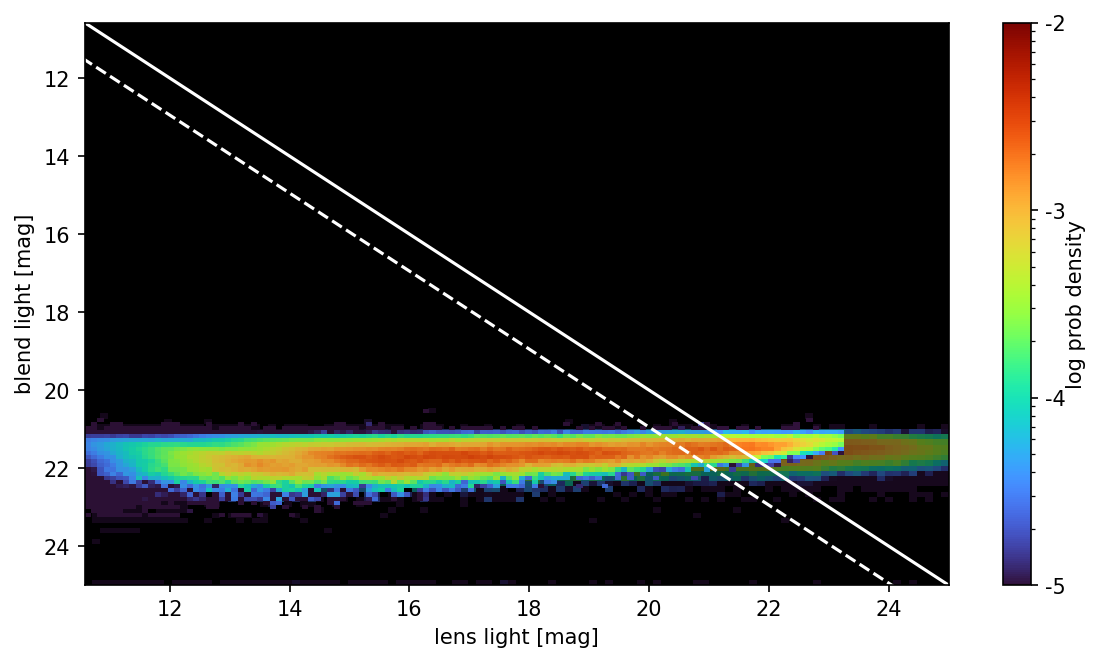}
\end{subfigure}
\caption{DLC results for Gaia21azb. Top: Mass-distance plot. Bottom: Blend-lens plot.}
\label{fig:Gaia21azb DLC}
\end{figure}

The light curve of Gaia21azb is shown in Figure \ref{fig:Gaia21azb_lc}, and its results for the DLC analysis are in Figure \ref{fig:Gaia21azb DLC}. Its higher distance limit is $\sim 3.7$ kpc, and the lens lower mass limit is $\sim$0.5 $M_{\odot}$.

The median mass derived from the DLC analysis, considering the Einstein radius limit, is $\sim1.0$ $\msun$. The dark lens probability for Gaia21azb falls within the range of 86-95\%. Additionally, we estimated the transverse velocity of the lens to exceed 150 km s$^{-1}$, which lies between the typical velocities of recycled and ordinary pulsars reported by \cite{Hobbs:2005}. Based on these findings, we conclude that the lens is most likely dark and may plausibly be either a high-mass white dwarf or a low-mass neutron star.

\subsection{GaiaDR3-ULENS-067} \label{sec:GaiaDR3-ULENS-067}

\begin{figure}[htbp]
    \centering
    \includegraphics[width=8.5cm]{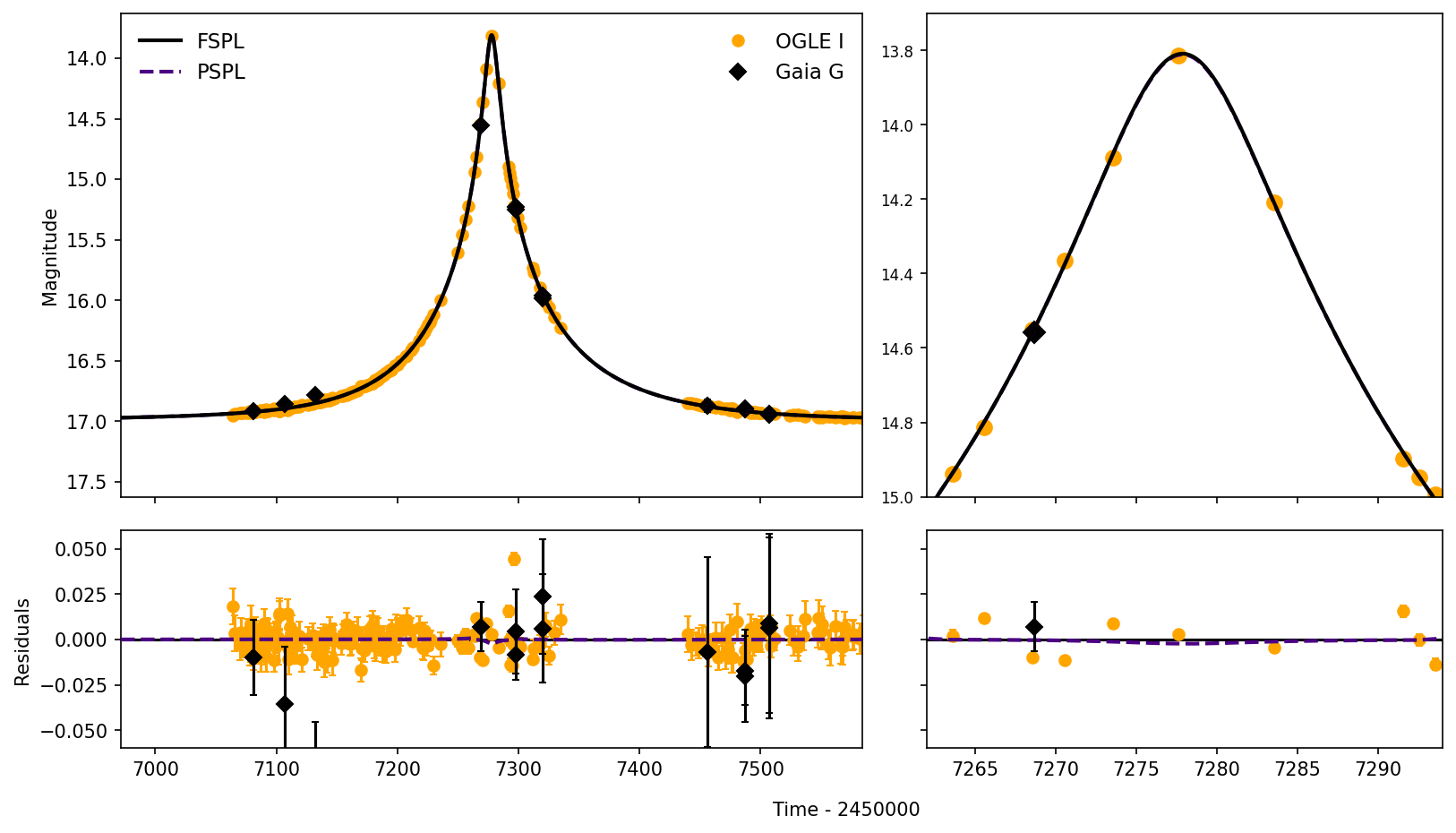}
    \caption{Light curve of GaiaDR3-ULENS-067 with the data from Gaia and OGLE. The solid line shows the parallax FS model, while the dashed line shows a parallax model for the point source. The right figure shows the zoom-in on the peak of the light curve.}
    \label{fig:GaiaDR3-ULENS-067_lc}
\end{figure}

\begin{figure}
    \centering
    \includegraphics[width=8.5cm]{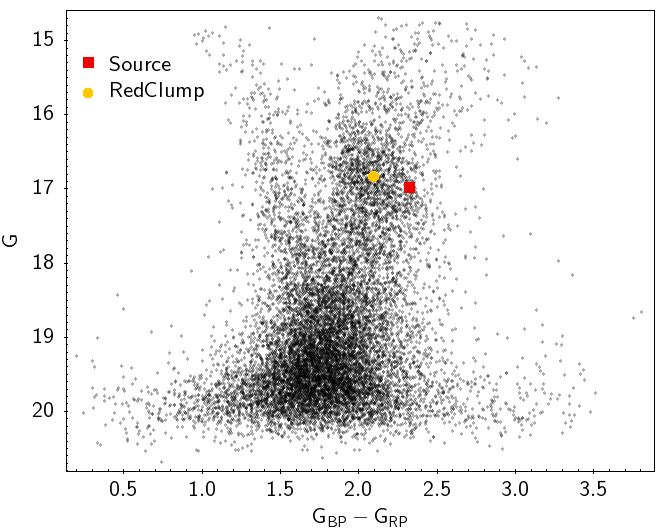}
    \caption{Colour-magnitude diagram (CMD) for GaiaDR3-ULENS-067. The red square represents the source, the yellow circle marks the RC centroid, and the black dots denote objects within a $5'$ radius of the source.}
    \label{ULENS-067_CMD}
\end{figure}

\begin{figure}[h]
\centering
\begin{subfigure}
  \centering
  \includegraphics[width=9cm]{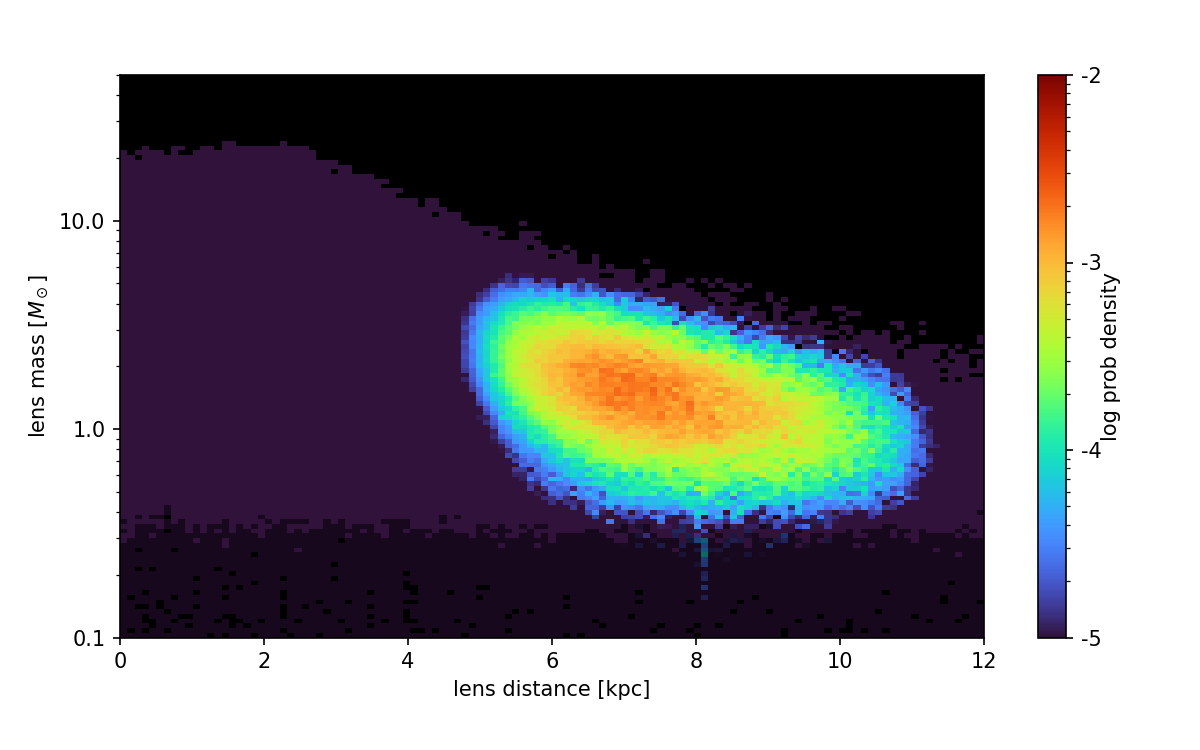}
\end{subfigure}

\begin{subfigure}
  \centering
  \includegraphics[width=8.33cm]{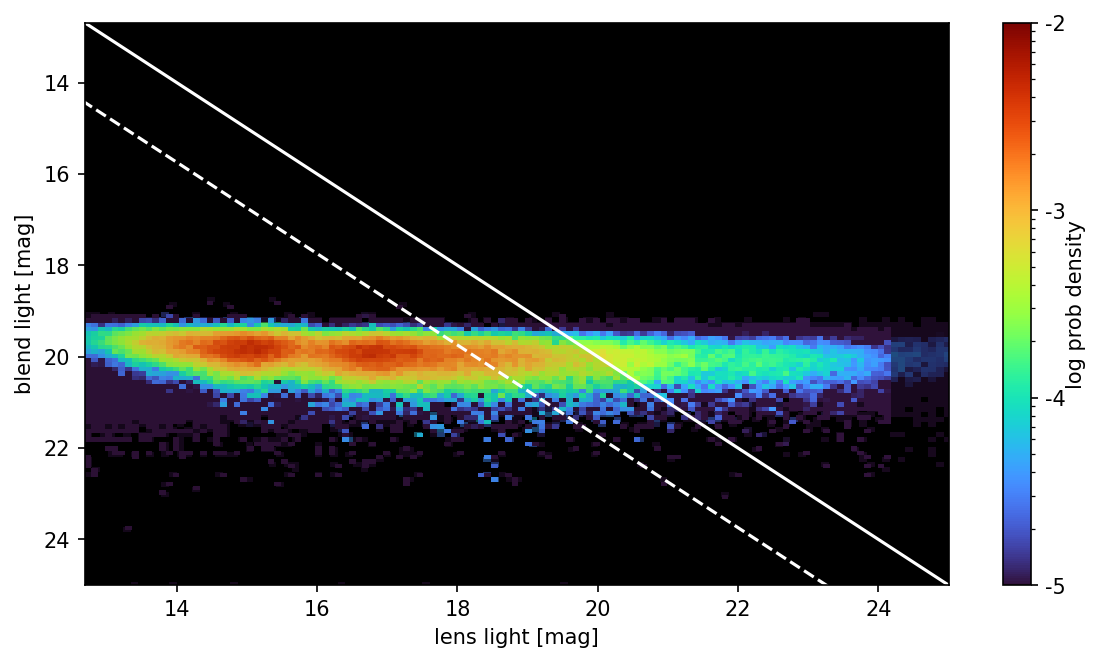}
\end{subfigure}
\caption{DLC results for GaiaDR3-ULENS-067. Top: Mass-distance plot. Bottom: Blend-lens plot.}
\label{fig:GaiaDR3-ULENS-067 DLC}
\end{figure}

Based on the CMD, the source in GaiaDR3-ULENS-067 is identified as a RC star from the Galactic Bulge (Figure \ref{ULENS-067_CMD}). Consequently, we estimate the distance to it to be 8$\pm2$ kpc \citep{Zoccali2016}. To derive the dereddened colour of the source, crucial for determining its angular radius, we employed the method outlined in \cite{Yoo2004}. Specifically, we utilised the following formula:

\begin{equation}
    ( (G_\mathrm{BP}-G_\mathrm{RP}), G)_0 = ( (G_\mathrm{BP}-G_\mathrm{RP}), G)_\mathrm{RC, 0} + \Delta ( (G_\mathrm{BP}-G_\mathrm{RP}), G)
.\end{equation}

In this equation, $( (G_\mathrm{BP}-G_\mathrm{RP}), G)_0$ and $( (G_\mathrm{BP}-G_\mathrm{RP}), G)_\mathrm{RC, 0}$ represent the dereddened colours and apparent magnitudes of the source and the centroid of the RC, respectively. $\Delta ( (G_\mathrm{BP}-G_\mathrm{RP}), G)$ denotes the difference between the measured source and RC centroid parameters extracted from the CMD. For the source they are equal to $( (G_\mathrm{BP}-G_\mathrm{RP}), G)=(2.32,16.98\pm0.02)$\footnote{In this and similar notations, magnitudes are the implied units, but they are omitted for clarity.} and for the RC $( (G_\mathrm{BP}-G_\mathrm{RP}), G)_\mathrm{RC}=(2.10,16.83\pm0.03)$. 

We obtained the values for the colour $(G_\mathrm{BP}-G_\mathrm{RP})_\mathrm{RC, 0}$ and absolute magnitude, $M_\mathrm{G, RC, 0}$, of the RC centroid from \cite{RedClumpColor}. This paper provides $G_\mathrm{RC, 0}$ two values for distances: calculated using the inverted Gaia parallax and taken from \cite{Bailer-Jones2018, 2021Bailer-Jones}. Furthermore, it offers values for two RC populations: high-$\alpha$ and low-$\alpha$. Since the RC in the Bulge belongs to the high-$\alpha$ population, we calculated $G_\mathrm{RC,0}$ using the average over two distances corresponding to this population. The resulting values are ($(G_\mathrm{BP}-G_\mathrm{RP})$, $G$)$_\mathrm{RC, 0}$ = (1.21$\pm$0.07, 15.09$\pm$0.24).

Finally, combining all of the necessary values, we obtained $( (G_\mathrm{BP}-G_\mathrm{RP}), G)_0 = (1.43 \pm 0.072, 15.24 \pm 0.24)$. Now we could use $G_0$ to calculate extinction in the G filter as follows: $A_G=G-G_0=1.74 \pm 0.24$. Then we transformed the colour $(G_\mathrm{BP}-G_\mathrm{RP})_0$ to ($(V-I)$, $I$)$_0$ using relations from GDR3 documentation\footnote{\href{https://gea.esac.esa.int/archive/documentation/GDR3/Data_processing/chap_cu5pho/cu5pho_sec_photSystem/cu5pho_ssec_photRelations.html}{gea.esac.esa.int/archive/documentation/GDR3}} and calculated the source size with a relation for giants from \cite{Adams2018}.

The light curve for GaiaDR3-ULENS-067 is shown in Figure \ref{fig:GaiaDR3-ULENS-067_lc} and the DLC results are in Figure \ref{fig:GaiaDR3-ULENS-067 DLC}. The estimated mass limit of $0.3-0.4$ $\msun$, along with a dark lens probability exceeding 80\%. While this mass limit does not significantly affect the DLC result, the median mass derived from the DLC for the $u_0 > 0$ model is $\sim1.70$ $\msun$. This value exceeds the Chandrasekhar limit, indicating that a dark remnant with this mass would be a neutron star. The median mass for the $u_0 < 0$ model is slightly lower at $\sim1.18$ $\msun$, indicating that a high-mass white dwarf or low-mass neutron star could act as a lens. 

The transverse velocity estimate further supports the possibility that this lens is a neutron star. For $u_0 > 0$, the median transverse velocity of the lens is $\sim320$ km s$^{-1}$, and for $u_0 < 0$, it is $\sim300$ km s$^{-1}$. The median value is higher than the mean for the pulsar transverse velocity distribution reported by \cite{Hobbs:2005}.

Our results are in good agreement with \cite{Rybicki2024}. They analyse this event (listed as OGLE-2015-BLG-0149) using a similar method and incorporating Spitzer data. They obtain a similar mass estimation and also point out that this lens as a strong stellar remnant candidate.

\subsection{Gaia21dpb} \label{sec:Gaia21dpb}
\begin{figure}[!htbp]
    \centering
    \includegraphics[width=8.5cm]{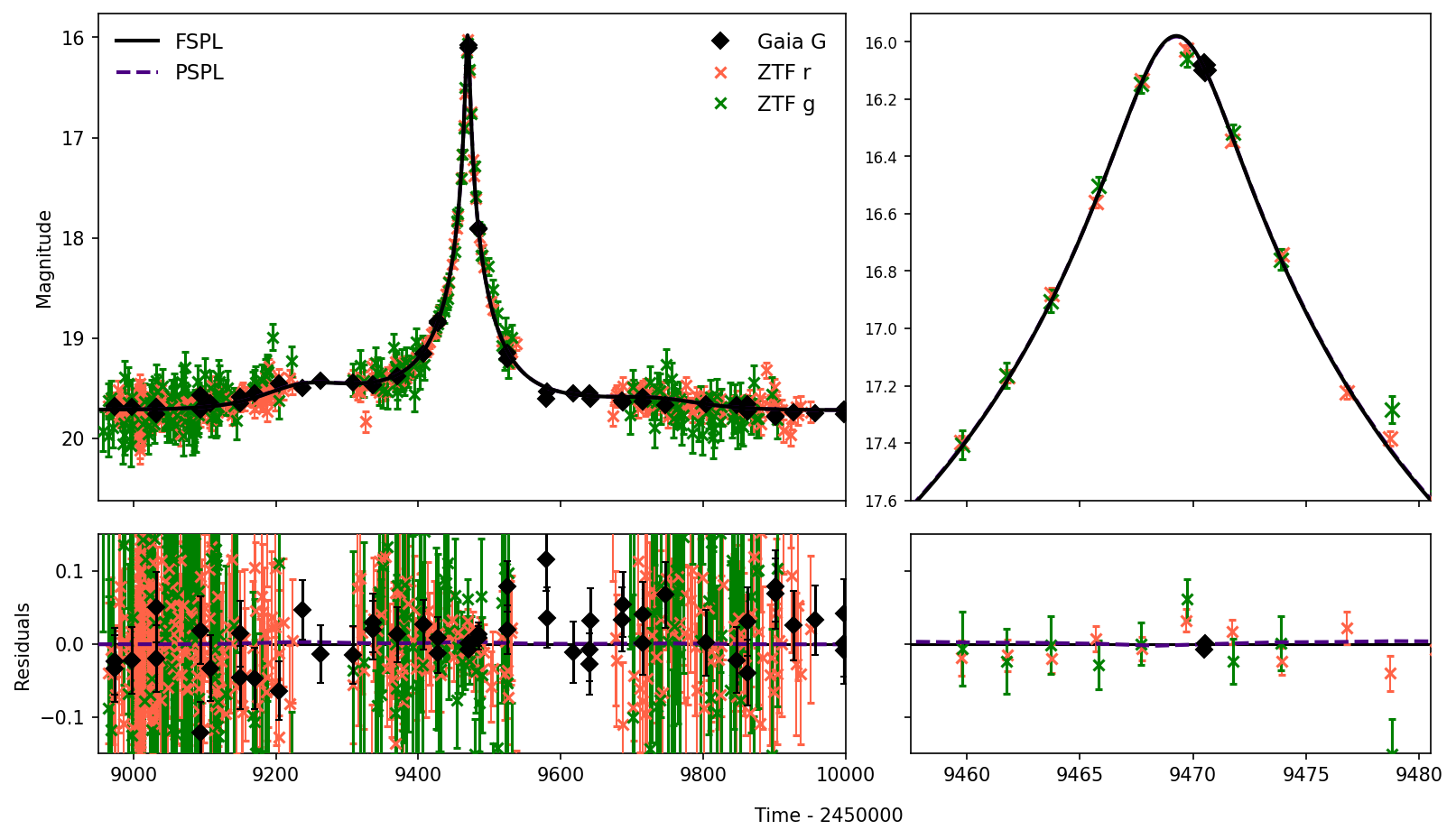}
    \caption{Light curve of Gaia21dpb with the data from Gaia and ZTF. The solid line shows the parallax FS model, while the dashed line shows a parallax model for the point source. The right figure shows the zoom-in on the peak of the light curve.}
    \label{fig:Gaia21dpb_lc}
\end{figure}

\begin{figure}[h]
\centering
\begin{subfigure}
  \centering
  \includegraphics[width=9cm]{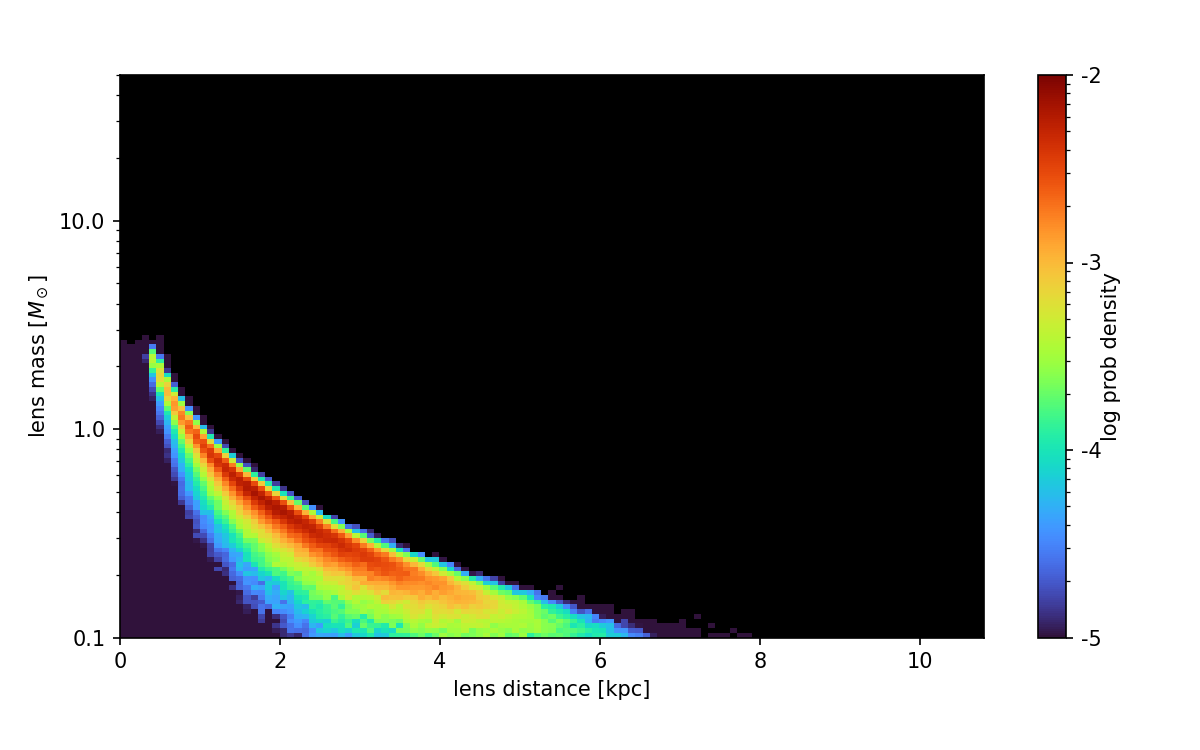}
\end{subfigure}

\begin{subfigure}
  \centering
  \includegraphics[width=8.33cm]{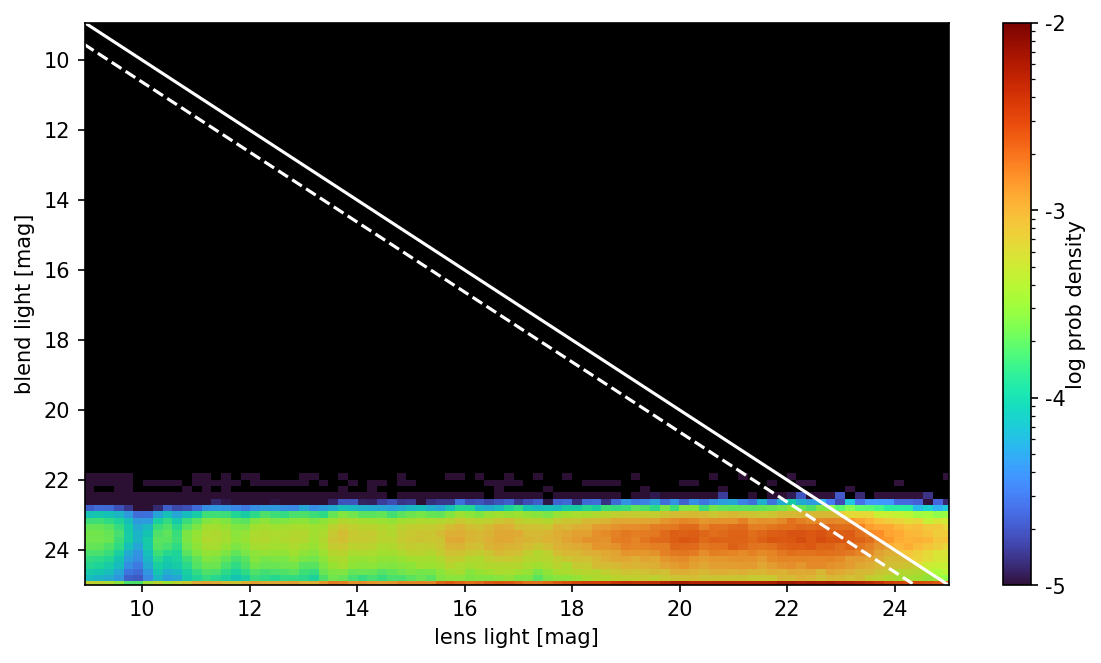}
\end{subfigure}
\caption{DLC results for Gaia21dpb. Top: Mass-distance plot. Bottom: Blend-lens plot.}
\label{fig:Gaia21dpb DLC}
\end{figure}

As the blending parameter was around 1.1 in the ZTF g filter, we fixed the blend flux in the ZTF g filter to 0 during the modelling. This made our model more physically plausible without significantly changing the results.

Based on the CMD, the source in Gaia21dpb is identified as a main-sequence star. Leveraging this information, we estimated the distance to the source. Initially, we converted the dereddened PS1 colours to UBVRI colours, specifically (V-I)$_0$ and (V-R)$_0$, as is described in Section \ref{sec:source star}. These values are (V-I)$_0=0.40^m\pm0.02$ and (V-R)$_0=0.15^m\pm0.02$. Next, to determine the star type, we compared these colours with the values provided by \cite{PecautMamajek2013}. According to this comparison, the source is classified as approximately F0V to F1V type. By averaging the absolute magnitudes of these types, we obtained a value of $M_V = 2.665^m$. \cite{PecautMamajek2013} did not provide errors for the absolute magnitude, so we added a conservative error of 0.5$^m$, similar to \cite{Kruszynska2022}. We then calculated the distance using the well-known formula

\begin{equation}
    D_\mathrm{S} = 10^{\frac{V_0-M_\mathrm{V}+5}{5}}
.\end{equation}

The $V_0$ value was also derived from dereddened PS1 photometry and is equal to $17.34^m\pm0.01$. The calculated distance falls within the range of 6.84 to 10.83 kpc, with a central value of 8.61 kpc, which is subsequently utilised in the DLC procedure.

The light curve of Gaia21dpb is demonstrated in Figure \ref{fig:Gaia21dpb_lc}, while the DLC results are in Figure \ref{fig:Gaia21dpb DLC}. The lower mass limit for this event is below $\sim0.014$ $M_{\odot}$, providing limited information about its mass or nature. However, the DLC results still suggest the possibility of a low-mass white dwarf, since its mass is $\sim0.4$ $\msun$. Furthermore, its probability of being dark falls within the range of 80-90\%, indicating a high likelihood of it being a dark object.

\section{Discussion} \label{sec:discussion}

In this work, we have analysed four unusual microlensing events: Gaia21efs, Gaia21azb, GaiaDR3-ULENS-067, and Gaia21dpb. Their light curves are moderately or highly magnified, but they do not exhibit any distinct FS effect. We constrained this effect and obtained the lower limits for the angular Einstein radius, and hence for the lens mass for each event. One of the possible explanations for the absence of the FS effect, which motivated us to pursue this work, is a large Einstein radius, as it is inversely proportional to the parameter, $\rho$, measuring the FS effect. This is promising in the context of searches for massive stellar remnants, as a large Einstein radius typically implies a large lens mass.

We find that the lenses in Gaia21efs, Gaia21azb, and GaiaDR3-ULENS-067 are the stellar remnant candidates, while Gaia21dpb presents a slightly weaker case. We make this statement based on several reasons. First, the lower mass limits for the first three events are $>0.2$ $M_{\odot}$, which is roughly the lowest possible mass for a white dwarf. Second, the blending is close to unity in these events, implying a dark lens. Finally, the lens light distribution analysis from the DLC procedure favours the hypothesis that these are indeed dark lenses rather than faint low-mass main-sequence stars.

In terms of particular remnant types, the lens in Gaia21efs is most likely a massive white dwarf; however, the estimate for its transverse velocity is relatively high, and could indicate the presence of a neutron star. Gaia21azb presents a similar case. Although a white dwarf is also suggested by the DLC analysis for Gaia21dpb, the evidence is not as strong due to the low mass limit. In GaiaDR3-ULENS-067, the large estimated mass and high transverse velocity of the lens strongly suggest that it is a neutron star. However, the possibility that it is a massive white dwarf cannot be ruled out.

There are other works in which dark isolated stellar remnant candidates are identified using microlensing light curves from Gaia (Section \ref{sec:introduction}). Along with these candidates, the four lenses found in this work may contribute to the sample of isolated remnants in the Galaxy, allowing us to study their mass and spatial distribution.

As is described in Section \ref{sec:introduction}, various methods exist for measuring lens masses, but they often present technical challenges or are applicable only to specific, uncommon cases. In this study, we focus on constraining the lens mass, which, as we demonstrate, can provide valuable insights into the lens nature. Our methodology is straightforward,  applicable to events with $u_0 \lesssim 0.1$, and exceptionally high magnification is not required for obtaining meaningful results. Applying this methodology to a larger set of events could be a valuable direction for future research.

Similar approaches were mentioned in Section \ref{sec:introduction} \citep{Smith2002, Shvartzvald2014, Zang2018_FS, Bachelet2022}. These studies determine $\rho_\mathrm{lim}$, and the latter three use techniques analogous to the DLC procedure. Although they share similarities with our work, we apply our methodology to single-lens, single-source events as a systematic approach to identifying stellar remnant candidates.

A primary limitation of our approach is the difficulty in predicting whether the Einstein radius limit will be meaningful from initial observations. Analysing our four events along with those from the four cited papers, we find no clear distinguishing characteristics. While magnification can be a factor, it appears that the method can be applied to moderately magnified events. For instance, Gaia21efs, the event with the highest Einstein radius and mass limit, has a maximum magnification of $\sim28$. In contrast, Gaia21azb, the most highly magnified event with a maximum magnification of $\sim80$, does not exhibit a particularly high mass limit. However, data quality near the light curve peak is crucial for accurately measuring $\rho$. Therefore, while there may be numerous sufficiently magnified events, the method may be ineffective if data are limited or have high uncertainties.

When searching for massive objects, it is advisable to assess the source radius beforehand to ensure it is not excessively small. It is generally more advantageous to plan follow-up observations with a larger source radius. With a smaller Einstein radius, the FS effect will probably be observed, enabling the lens mass to be derived, although it might not be significant. Otherwise, there is a higher chance of a substantial mass limit. This strategy may create promising prospects for faint yet highly magnified microlensing events discovered by surveys such as LSST, which could then be followed up using smaller telescopes.

Additionally, other parameters such as source angular radius, distance, and extinction are required, and in the case of a lack of additional measurements, empirical methods may be necessary to estimate these values. Thus, future improvements to our results may stem from more precise measurements of the sources. Given the spectroscopic follow-up of numerous microlensing events, it is realistic to obtain distance (as for Gaia21efs) and/or extinction. While interferometry can measure the angular radius, this level of complexity would be excessive for our purposes, especially considering our inability to measure $\rho$ and derive the lens mass. 

It is worth noting that the choice of the 95\% lower credible interval for defining $\theta_{\rm E,lim}$ is not unique and represents a methodological choice rather than a strict standard. Different studies adopt different thresholds depending on their goals: for example, \citet{Smith2002} use $2\sigma$, \citet{Shvartzvald2014} and \citet{Zhu2016} adopt $3\sigma$, and \citet{Bachelet2022} use a more conservative $10\sigma$ limit. A stricter threshold may be justified when the outcome influences costly follow-up observations. In our case, however, all data have already been collected, and the main purpose is to prioritise the analysis of potentially interesting events. We therefore opt for the 95\% credible interval as a practical compromise between strictness and informativeness. However, even when adopting a stricter $5\sigma$ threshold, the lower mass limits for Gaia21efs and Gaia21azb remain above 0.2 $\msun$. The limit for GaiaDR3-ULENS-067 is somewhat lower under this criterion, but its DLC results (high mass, dark lens probability, and transverse velocity) still strongly support its classification as a stellar remnant.

The compact nature of the candidate lenses can be most accurately validated through astrometric counterparts of the analysed photometric microlensing events. The astrometric series from Gaia Data Release 4 will encompass data from the first 5.5 years of the mission; therefore, only the GaiaDR3-ULENS-067 event (from 2015) can be analysed using these data. However, it will not be publicly available until mid-2026. The remaining events, occurring in 2021, can be properly analysed with Gaia Data Release 5, the full Gaia catalogue, estimated for release in 2030. However, some astrometric shifts might be detectable even in the Gaia Data Release 4 data. The accuracy of the mass measured using these data depends on factors such as the source brightness and lens mass. The most precise mass measurements are expected for bright sources lensed by massive objects, assuming average Gaia coverage. Therefore, Gaia21efs with a G-band magnitude range of 15.7$^m$ to 12.1$^m$, has promising potential for accurate mass measurement.

Apart from astrometric microlensing, imaging can also be used to verify the nature of the lenses. For example, young neutron stars would be detectable in X-rays and young white dwarfs in ultraviolet, provided they are relatively nearby. Therefore, there is a chance of detecting lenses in Gaia21efs or Gaia21dpb, which are situated at a distance of $\sim$ 1.8 kpc. Conversely, since low-mass main-sequence stars emit primarily in the infrared, the lack of lens detection in this band can decisively rule out the presence of such a star. This absence would strengthen the case for the lens being a compact object, potentially with a high temperature and emitting mainly in ultraviolet or X-rays. A similar approach has previously been used, as is demonstrated in \cite{Blackman2021}.

Future surveys could benefit from an adaptation of the methodology used in this study. For instance, the Rubin Observatory is anticipated to observe thousands of microlensing events with dense temporal coverage during its operation \citep{Sajadian2019}. If the analysis of light curves similar to those analysed in this work produces significant mass limits, it will provide a strong argument for conducting follow-up astrometric time-series observations of these events, such as with the Roman Space Telescope, thereby enabling a more efficient use of observational resources.

\section{Conclusions} \label{sec:conclusions}

Gaia21efs, Gaia21azb, GaiaDR3-ULENS-067, and Gaia21dpb are microlensing events with moderately or highly magnified light curves. Despite this, they lack the FS effect, potentially due to the large Einstein radius, and hence also possibly due to the high mass of the lens. For these events, we derived lower limits for their angular Einstein radii and lens masses by constraining the $\rho$ distribution.

Our methodology involved three key steps: modelling the light curve, determining source parameters, and conducting the DLC analysis. It is most effective for events with $u_0<0.1$ and sufficient observational data. Given the challenges associated with lens mass measurement methods, this approach provides a reliable estimate while remaining relatively simple. Therefore, it can be incorporated into follow-up optimisation efforts for large numbers of microlensing events to be detected by facilities such as Rubin and Roman.

Gaia21efs, Gaia21azb, and GaiaDR3-ULENS-067 are estimated to have lower mass limits in the range of $0.3-0.8$ $\msun$. In contrast, the mass limit for Gaia21dpb is significantly lower, at $\sim0.014$ $\msun$, which minimally contributes to this analysis. Considering the mass limits and the DLC analysis, we conclude that the lenses in Gaia21efs and Gaia21azb are most likely white dwarfs with masses of $\sim1$ $\msun$, although their relatively high transverse velocities leave open the possibility that they could be neutron stars. Gaia21dpb is likely a white dwarf with a mass of $\sim 0.4$ $\msun$. The lens in GaiaDR3-ULENS-067 can be a neutron star, with a mass estimated between $\sim 1.18-1.70$ $\msun$ depending on the model, and a high transverse velocity of $\sim300-320$ km s$^{-1}$. The nature of lenses can be confirmed by measuring their masses using astrometric data from Gaia or using imaging techniques.

If confirmed, these lenses, together with stellar remnant candidates identified in other studies, could significantly improve our knowledge of the mass and spatial distribution of dark remnants. In turn, this could provide valuable insights into the evolution of massive stars and binary systems, shedding light on how supernovae have shaped the history of the Milky Way.

\bibliography{bibs}
\bibliographystyle{aa}

\begin{appendix}

\begin{landscape}
\section{List of BHTOM facilities} \label{app:BHTOM facilities}

The following table lists BHTOM facilities which collected data for this study.

\begin{table}[htbp]

\centering

\caption{Facilities of BHTOM which collected data for analysed events.}
\small
\begin{tabular}{llrlrrr}

\hline
 Name in BHTOM               & Telescope                        &   Size [m] & Location                                                                    &        Lon [$^{\circ}$] &      Lat [$^{\circ}$] &   Alt [m] \\ 
 \hline
 \hline
 ASV1.4\_Andor              & ASV Telescope Milanković         &     1.4   & Astronomical Observatory, Belgrade, Serbia                                       &   21.5556  &  43.1402 &  1150 \\
 Adyu60\_Andor-934          & ASA DM160                        &     0.6   & Adiyaman Observatory, Turkey                                                &   38.2254  &  37.7517 &   600 \\
 BIALKOW\_ANDOR-DW432       & Białków 0.6-m    &     0.6   & Astronomical Institute, University of Wroclaw, Poland                       &   16.6578  &  51.4742 &   150\\
 C2PU\_QHY600M              & C2PU Omicron                     &     1.04  & Cote Azur Observatory, France                                               &    6.92301 &  43.7537 &  1260 \\
 Flarestar-MPC171\_G2-1600  & Meade SSC-10                     &     0.25  & Flarestar Observatory, Malta                                                &   14.4697  &  35.9102 &   126 \\
 GeoNAO\_SXVR-H36           & GeoNAO SCTelescope        &     0.355 & Georgian National Astrophysical Observatory, Georgia                        &   42.8194  &  41.7544 &  1650 \\
 HAO68\_G2-1600             & Horten 0.68-m                    &     0.675 & Horten Local Observatory, Norway                                            &   10.4368  &  59.4171 &    86 \\
 IST60\_Andor-888           & IST60 RC Lightweight             &     0.6   & Ulupinar Observatory, Turkey                                                &   26.4753  &  40.0993 &   410 \\
 LCOGT-CTIO-1m\_4K          & LCOGT\_CTIO100                    &     1.0     & Cerro Tololo Inter-American Observatory, Chile                              &  -70.8048  & -30.1674 &  2201 \\
 LCOGT-HO-2m\_Spectral      & Faulkes Telescope North          &     2.0     & Haleakala High Altitude Observatory, USA                          & -156.258   &  20.707  &  3034 \\
 LOIANO1.52\_BFOSC          & LOIANO1.52                       &     1.52  & Loiano Observatory, Italy                                                   &   11.3344  &  44.2591 &   785 \\
 MOLETAI-35cm\_CCD4710      & 35-cm Maksutov                   &     0.35  & Moletai Observatory, Lithuania                                              &   25.5631  &  55.3159 &   200 \\
 OAUJ-CDK500\_F42           & OAUJ-CDK500                      &     0.5   & Jagiellonian University Astronomical Obs., Poland            &   19.8245  &  50.054  &   318 \\
 OAUJ-CDK500\_U47           & OAUJ-CDK500                      &     0.5   & Jagiellonian University Astronomical Obs., Poland            &   19.8245  &  50.054  &   318 \\
 Ondrejov50                & Richi-Cretien 0.65m                                &     0.65     & Ondřejov Observatory, Czech Republic                                                                    &    14.781       &   49.909     & 521\\
 PIWNICE90\_C4-16000EC & Schmidt-Cassegrain 90-cm & 0.9 & Nicolaus Copernicus University, Piwnice, Poland &	18.556 & 53.094 & 100\\
 RRRT\_SBIG-STX16803        & RRRT                             &     0.6   & Fan Mountains Observatory, United States                                    &  -78.6942  &  37.8791 &  1683 \\
 SUTO-Otivar\_ASI1600MM     & SUTO-Otivar                      &     0.3   & Silesian University of Technology Obs., Spain                      &   -3.70316 &  36.8282 &   200 \\
 SUTO-Pyskowice\_ATIK11000M & SUTO-Pyskowice                   &     0.3   & Silesian University of Technology Obs., Poland                     &   18.714   &  50.3982 &   240 \\
 T100\_SI1100               & T100                             &     1.0     & Tubitak Observatory, Turkey                                                 &   30.3356  &  36.8242 &  2500 \\
 TERSKOL-60\_SBIG-STL-1001     & 60-cm Zeiss Cassegrain telescope &     0.6    & Terskol Observatory, Russian Federation                                     &   42.4994  &  43.2764 &  3143 \\
 TJO\_MEIA2                 & Telescope Joan Oró     &     0.8   & The Montsec Astronomical Observatory, Spain                                 &    0.72969 &  42.0517 &  1570 \\
 TRT-SBO-0.7\_Andor-934     & TRT-SBO                          &     0.7   & NARIT, Australia             &  149.081   & -31.2817 &  1020 \\
 TRT-SRO-0.7\_Andor-934     & TRT-SRO                          &     0.7   & NARIT, USA         & -119.413   &  37.0703 &  1405 \\
 Tomo-e Gozen 1.05 Kiso    & Kiso Schmidt telescope                                &     1.05     & Kiso Observatory, University of Tokyo, Japan                                                                    &     137.626      &     35.797   &     1130  \\
 VATT\_Vatt4k               & VATT                             &     1.83  & Vatican Observatory, Mount Graham, USA & -109.892   &  32.7013 &  3191 \\
 ZAO\_G2-1600               & 20-cm SCT Telescope              &     0.2   & Znith Astronomy Observatory, Malta                                          &   14.44    &  35.9204 &   113 \\
 pt5m                      & Officina Stellare Ritchey-Chretien                                &     0.5     &  Roque de los Muchachos Observatory, Spain                                                                   &        17.885   &     28.757   &     2396\\ \hline
\end{tabular}
\label{tab:telescopes}
\tablefoot{Columns: name of a telescope in BHTOM, telescope details, size, location, longtitude, latitude, altitude.}
\end{table}
\end{landscape}

\onecolumn
\section{Photometric data} \label{app:photometric data}

The following tables present all the photometric data available for each event, detailing the number of data points per facility before and after cleaning. The tables also include the minimum and maximum MJD for the cleaned data. Facilities whose data was used in the analysis are highlighted in bold, as are the filters applied. A superscript $^B$ indicates that data from a facility was processed using BHTOM.

\begin{table*}[hbt!]
\caption{Photometric data for Gaia21azb.}
    \centering
    \begin{tabular}{rrrrrr}
    \hline
Facility	&	Filter	&	Before cleaning	&	After cleaning	&	Min MJD	&	Max MJD	\\
\hline
\hline
\textbf{Gaia Alerts}	&	\textbf{G}	&	163	&	163	&	57077.68   	&	60264.83	\\
\textbf{ZTF}	&	\textbf{ZTF g, r}	&	114	&	77	&	58277.46   	&	59750.42	\\
ATLAS	&	ATLAS o, c, H	&	749	&	0	&	-	&	-	\\
PS1	&	PS1 g, r, i, z	&	4	&	0	&	-	&	-	\\
2MASS	&	2MASS J, H, K	&	3	&	0	&	-	&	-	\\
NEOWISE	&	W1, W2	&	58	&	0	&	-	&	-	\\
\textbf{LCOGT1m$^B$}	&	\textbf{g, i}	&	685	&	663	&	59280.51   	&	59440.17	\\
 BIALKOW\_ANDOR-DW432$^B$	&	B,V,R,I,\textbf{g}	&	200	&	7	&	59342.04   	&	59342.07	\\
\textbf{ASV1.4\_Andor$^B$}	&	B, V, \textbf{R, I}, g	&	166	&	47	&	59345.00   	&	59463.81	\\
\textbf{Ondrejov50$^B$}	&	B, V, R, U, I, g, \textbf{r}, i	&	113	&	13	&	59353.94   	&	59511.78	\\
 \textbf{TRT-SBO-0.7$^B$}	&	V, \textbf{R,I}	&	79	&	33	&	59347.67   	&	59372.70	\\
\textbf{LCOGT-HO-2m\_Spectral$^B$}	&	\textbf{g,r,i,}z	&	47	&	23	&	59386.33   	&	59454.46	\\
MOLETAI-35cm\_CCD4710$^B$	&	V,R,I,i	&	30	&	0	&	-	&	-	\\
\textbf{Adyu60\_Andor-934$^B$}	&	g,\textbf{r},i,u	&	17	&	7	&	59338.92   	&	59358.90	\\
\textbf{TJO\_MEIA2$^B$}	&	V, \textbf{R,I}	&	16	&	12	&	59345.04   	&	59447.00	\\
Tomo-e Gozen 1.05 Kiso$^B$	&	R,r	&	14	&	0	&	-	&	-	\\
\textbf{T100\_SI1100$^B$}	&	\textbf{r,}i	&	4	&	2	&	59395.04   	&	59395.05	\\
\textbf{pt5m$^B$}	&	\textbf{R}	&	1	&	1	&	59290.23   	&	59290.23 	\\
\hline
    \end{tabular}
    \label{tab:obs-Gaia21azb}
\end{table*}

\begin{table*}[hbt!]
\caption{Photometric data for GaiaDR3-ULENS-067.}
    \centering
    \begin{tabular}{rrrrrr}
\hline
Facility	&	Filters	&	Before cleaning	&	After cleaning	&	Min MJD	&	Max MJD	\\
\hline
\hline
\textbf{Gaia DR3}	&	\textbf{G}, G$_{BP}$, G$_{RP}$	&	53	&	44	&	56910.46  	&	57879.28	\\
\textbf{OGLE}	&	\textbf{OGLE I}	&	3169	&	3169	&	55260.38  	&	58923.34	\\
ATLAS	&	ATLAS o, c, i, g, r, H	&	2204	&	0	&	-	&	-	\\
2MASS	&	2MASS J, H, K	&	3	&	0	&	-	&	-	\\
ALLWISE	&	W1, W2	&	96	&	0	&	-	&	-	\\
NEOWISE	&	W1, W2	&	541	&	0	&	-	&	-	\\
DECAPS	&	DECAPS g, r, z	&	3	&	0	&	-	&	-	\\
\hline
    \end{tabular}
    \label{tab:obs-ULENS-067}
\end{table*}

\begin{table*}[hbt!]
\caption{Photometric data for Gaia21dpb.}
    \centering
    \begin{tabular}{rrrrrr}
\hline
Facility	&	Filters	&	Before cleaning	&	After cleaning	&	Min MJD	&	Max MJD	\\
\hline
\hline
\textbf{Gaia Alerts}	&	\textbf{G}	&	158	&	136	&	56864.38 	&	59997.43	\\
\textbf{ZTF}	&	\textbf{ZTF g, r,} i	&	3615	&	1712	&	58218.49 	&	59951.15	\\
NEOWISE	&	W1, W2	&	3	&	0	&	-	&	-	\\
PS1	&	PS1 g, r, i, z	&	4	&	0	&	-	&	-	\\
HAO68\_G2-1600$^B$ 	&	B, V, R, I, r	&	39	&	0	&	-	&	-	\\
LOIANO1.52\_BFOSC$^B$	&	V, R, I	&	18	&	0	&	-	&	-	\\
PIWNICE90\_C4-16000EC$^B$ & U, B, V, R, I & 16 & 0 & - & - \\
\hline
    \end{tabular}
    \label{tab:obs-Gaia21dpb}
\end{table*}

\twocolumn

\begin{table*}[hbt!]
\caption{Photometric data for Gaia21efs. }
    \centering
    \begin{tabular}{rrrrrr}
    \hline
Facility	&	Filter	&	Before cleaning	&	After cleaning	&	Min MJD	&	Max MJD	\\
\hline
\hline
\textbf{Gaia Alerts}	&	\textbf{G}	&	227	&	220	&	56962.36	&	60272.59	\\
\textbf{ZTF}	&	\textbf{ZTF g, r,} i	&	3350	&	1377	&	58198.52	&	60125.44	\\
\textbf{ATLAS}	&	\textbf{ATLAS o, c,} H	&	3215	&	2377	&	57302.32	&	60259.29	\\
2MASS	&	2MASS J,H,K	&	3	&	0	&	-	&	-	\\
ALLWISE	&	W1,W2	&	64	&	0	&	-	&	-	\\
NEOWISE	&	W1,W2	&	762	&	0	&	-	&	-	\\
PS1	&	PS1 g,r,i,z	&	4	&	0	&	-	&	-	\\
TRT-SRO-0.7\_Andor-934$^B$ 	&	U, V, R, I, u, g, r, i	&	2872	&	0	&	-	&	-	\\
\textbf{ BIALKOW\_ANDOR-DW432$^B$}	&	\textbf{B, V, R, I, r,} g	&	1019	&	224	&	59489.74	&	59518.86	\\
\textbf{LCOGT1m$^B$}	&	\textbf{g, i}	&	736	&	676	&	59481.84	&	59561.04	\\
\textbf{TJO\_MEIA2$^B$}	&	\textbf{B, V, R, I, r}	&	557	&	393	&	59519.86	&	59715.11	\\
\textbf{IST60\_Andor-888$^B$}	&	\textbf{V, R,} I, g, \textbf{r,} i	&	503	&	211	&	59517.69	&	59531.69	\\
\textbf{OAUJ\_Krakow-CDK500$^B$}	&	\textbf{B,V, R, I,} g, i, z	&	356	&	331	&	59514.77	&	60167.0	\\
\textbf{MOLETAI-35cm\_CCD4710$^B$}	&	\textbf{V, R, I, r}	&	185	&	166	&	59482.98	&	59709.0	\\
\textbf{GeoNAO\_SXVR-H36$^B$}	&	\textbf{V,} R, \textbf{I, r}	&	72	&	33	&	59522.68	&	59527.75	\\
\textbf{ RRRT\_SBIG-STX16803$^B$}	&	\textbf{V, R,} g, i	&	68	&	6	&	59696.28	&	59720.26	\\
\textbf{LOIANO1.52\_BFOSC$^B$}	&	\textbf{V, R, I, r}	&	63	&	49	&	59514.87	&	59555.75	\\
\textbf{VATT\_Vatt4k$^B$}	&	g, \textbf{r,} i	&	59	&	18	&	59522.15	&	59575.13	\\
OAUJ-CDK500\_F42$^B$	&	V, R, I	&	41	&	0	&	-	&	-	\\
\textbf{TERSKOL-60\_SBIG-STL-1001$^B$}	&	\textbf{V, R, I,} g, z	&	37	&	26	&	59514.67	&	59562.75	\\
\textbf{HAO68\_G2-1600$^B$}	&	\textbf{B, V, R, I}	&	31	&	23	&	59540.77	&	59674.06	\\
\textbf{Flarestar-MPC171\_G2-1600$^B$}	&	\textbf{V, I}	&	19	&	18	&	59520.71	&	59583.71	\\
\textbf{SUTO-Otivar\_ASI1600MM$^B$}	&	B, \textbf{V, I, r,} i	&	19	&	11	&	59700.09	&	59706.18	\\
C2PU\_QHY600M$^B$	&	i	&	10	&	0	&	-	&	-	\\
\textbf{ASV1.4\_Andor$^B$} 	&	\textbf{B, V, R, I}	&	8	&	8	&	59518.72	&	59518.72	\\
\textbf{SUTO-Pyskowice\_ATIK11000M$^B$}	&	B, \textbf{V, r,} i	&	6	&	4	&	59711.04	&	59711.06	\\
\textbf{ZAO\_G2-1600$^B$}	&	\textbf{R, I}	&	5	&	5	&	59569.76	&	59569.78	\\
\hline
    \end{tabular}
    \label{tab:obs-Gaia21efs}
\end{table*}

\FloatBarrier

\section{Photometric curve parameters} \label{app:curve-params}

The following tables present the fitted parameters.

\begin{table}[!htbp]
\centering
\begin{scriptsize}
\centering
\caption{Photometric model parameters for Gaia21dpb. An asterisk indicates that $u_0$ value is calculated using the mode.}
\footnotesize
\label{tab:Gaia21dpb-photometry}
\begin{tabular}{lll}
\hline
Parameter & $u_0>0^*$ & $u_0<0^*$ \\ 
\hline
\hline
        $t_{0,par}-2450000$ [d]	&	9469	&	9469	\\
        $t_0-2450000$ [d]	&	$9469.26^{+0.03}_{-0.03}$	&	$9469.27^{+0.03}_{-0.04}$	\\
        $u_0$	&	$0.0314^{+0.0010}_{-0.0007}$	&	$-0.0314^{+0.0007}_{-0.0011}$	\\
        $t_E$ [d]	&	$82.89^{+1.06}_{-1.04}$	&	$84.51^{+1.08}_{-1.04}$	\\
        $\pi_{EN}$	&	$-0.2585^{+0.0085}_{-0.0082}$	&	$-0.2529^{+0.0083}_{-0.0078}$	\\
        $\pi_{EE}$	&	$-0.2446^{+0.0081}_{-0.0081}$	&	$-0.2375^{+0.0078}_{-0.0078}$	\\
        $\rho_{lim}$	&	0.0249	&	0.0253	\\
        G$_0$ [mag]	&	$19.738^{+0.001}_{-0.001}$	&	$19.728^{+0.001}_{-0.001}$	\\
        $f_{s,G}$	&	$0.981^{+0.016}_{-0.016}$	&	$0.980^{+0.016}_{-0.016}$	\\
        ZTF r$_0$ [mag]	&	$19.726^{+0.001}_{-0.001}$	&	$19.726^{+0.001}_{-0.001}$	\\
        $f_{s, ZTF r}$	&	$0.998^{+0.014}_{-0.014}$	&	$0.998^{+0.015}_{-0.014}$	\\
        ZTF g$_0$ [mag]	&	$21.019^{+0.008}_{-0.007}$	&	$21.019^{+0.008}_{-0.007}$	\\
        $f_{s, ZTF g}$	&	1.0 (fixed)	&	1.0 (fixed)	\\
        $\chi^2$	&	1820	&	1820	\\
        $\frac{\chi^2}{dof}$	&	0.99	&	0.99	\\
\hline

\end{tabular}
\end{scriptsize}
\end{table}

\begin{table}[!htbp]
\centering
\begin{scriptsize}
\centering
\caption{Photometric model parameters for Gaia21azb. An asterisk indicates that $u_0$ value is calculated using the mode.}
\footnotesize
\label{tab:Gaia21azb-photometry}
\begin{tabular}{lll}
\hline
Parameter & $u_0>0^*$ & $u_0<0$ \\ 
\hline
\hline
$t_{0,par}-2450000$ [d]	&	9340.67	&	9340.67	\\
$t_0-2450000$ [d]	&	$9340.672^{+0.001}_{-0.001}$	&	$9340.672^{+0.001}_{-0.001}$	\\
$u_0$	&	$0.0124^{+0.0002}_{-0.0001}$	&	$-0.0124^{+0.0002}_{-0.0002}$	\\
$t_E$ [d]	&	$88.31^{+1.03}_{-0.96}$	&	$88.53^{+1.08}_{-1.03}$	\\
$\pi_{EN}$	&	$0.1448^{+0.0542}_{-0.0443}$	&	$0.1315^{+0.0473}_{-0.0418}$	\\
$\pi_{EE}$	&	$-0.0126^{+0.0162}_{-0.0210}$	&	$-0.0091^{+0.0147}_{-0.0193}$	\\
$\rho_{lim}$	&	0.0058	&	0.0057	\\
G$_0$ [mag]	&	$18.424^{+0.002}_{-0.001}$	&	$18.424^{+0.002}_{-0.001}$	\\
$f_{s,G}$	&	$0.942^{+0.013}_{-0.013}$	&	$0.944^{+0.012}_{-0.012}$	\\
ZTF g$_0$ [mag]	&	$20.164^{+0.004}_{-0.004}$	&	$20.164^{+0.004}_{-0.003}$	\\
$f_{s, ZTF g}$	&	$0.904^{+0.015}_{-0.015}$	&	$0.905^{+0.015}_{-0.014}$	\\
ZTF r$_0$ [mag]	&	$18.473^{+0.007}_{-0.006}$	&	$18.472^{+0.006}_{-0.006}$	\\
$f_{s, ZTF r}$	&	$0.990^{+0.012}_{-0.011}$	&	$0.990^{+0.012}_{-0.011}$	\\
i$_0$ [mag]	&	$17.842^{+0.013}_{-0.012}$	&	$17.840^{+0.012}_{-0.012}$	\\
$f_{s,i}$	&	$1.001^{+0.001}_{-0.002}$	&	$1.001^{+0.001}_{-0.002}$	\\
g$_0$ [mag]	&	$20.233^{+0.012}_{-0.011}$	&	$20.232^{+0.012}_{-0.011}$	\\
$f_{s,g}$	&	$0.920^{+0.002}_{-0.002}$	&	$0.920^{+0.002}_{-0.002}$	\\
R$_0$ [mag]	&	$18.350^{+0.014}_{-0.013}$	&	$18.349^{+0.014}_{-0.013}$	\\
$f_{s,R}$	&	$1.017^{+0.004}_{-0.009}$	&	$1.019^{+0.003}_{-0.009}$	\\
r$_0$ [mag]	&	$18.601^{+0.013}_{-0.012}$	&	$18.599^{+0.012}_{-0.011}$	\\
$f_{s,r}$	&	$0.918^{+0.005}_{-0.004}$	&	$0.917^{+0.004}_{-0.004}$	\\
I$_0$ [mag]	&	$17.354^{+0.014}_{-0.013}$	&	$17.352^{+0.013}_{-0.012}$	\\
$f_{s,I}$	&	$1.004^{+0.004}_{-0.005}$	&	$1.004^{+0.003}_{-0.005}$	\\
$\chi^2$	&	983	&	961	\\
$\frac{\chi^2}{dof}$	&	0.958	&	0.937 \\
\hline

\end{tabular}
\end{scriptsize}
\end{table}

\begin{table}[!htbp]
\centering
\begin{scriptsize}
\centering
\caption{Photometric model parameters for GaiaDR3-ULENS-067. An asterisk indicates that $u_0$ value is calculated using the mode.}
\footnotesize
\label{tab:GaiaDR3-photometry}
\begin{tabular}{lll}
\hline
Parameter & $u_0>0^*$ & $u_0<0^*$ \\ 
\hline
\hline
$t_{0,par}-2450000$ [d]	&	7277.87	&	7277.87	\\

$t_0-2450000$ [d]	&	$7277.87^{+0.01}_{-0.01}$	&	$7277.86^{+0.01}_{-0.01}$	\\

$u_0$	&	$0.0536^{+0.0007}_{-0.0005}$ 	&	$-0.0532^{+0.0003}_{-0.0006}$ \\

$t_E$ [d]	&	$102.08^{+0.47}_{-0.46}$	&	$101.35^{+0.37}_{-0.35}$	\\

$\pi_{EN}$	&	$0.0637^{+0.0176}_{-0.0169}$	&	$0.0874^{+0.0153}_{-0.0162}$	\\

$\pi_{EE}$	&	$-0.0469^{+0.0016}_{-0.0015}$	&	$-0.0469^{+0.0011}_{-0.0010}$	\\

$\rho_{lim}$	&	0.0258	&	0.0236	\\

G$_0$ [mag]	&	$16.9903^{+0.0003}_{-0.0003}$	&	$16.9902^{+0.0002}_{-0.0002}$	\\

$f_{s,G}$	&	$0.996^{+0.007}_{-0.007}$	&	$0.989^{+0.003}_{-0.004}$	\\

OGLE I$_0$ [mag]	&	$15.9217^{+0.0001}_{-0.0001}$	&	$15.9220^{+0.0001}_{-0.0001}$	\\

$f_{s, OGLE I}$	&	$0.976^{+0.007}_{-0.007}$ &	$0.970^{+0.004}_{-0.004}$	\\

$\chi^2$	&	1687	&  1688	\\
$\frac{\chi^2}{dof}$	&	0.995	&	0.998	\\
\hline

\end{tabular}
\end{scriptsize}
\end{table}

\begin{table}[!htbp]
\centering
\begin{scriptsize}
\centering
\caption{Photometric model parameters for Gaia21efs.}
\footnotesize
\label{tab:Gaia21efs-photometry}
\begin{tabular}{lll}
\hline
Parameter & $u_0>0^*$ & $u_0<0^*$ \\ 
\hline
\hline
$t_{0,par}-2450000$ [d]	&	9520.27	&	9520.27	\\
$t_0-2450000$ [d]	&	$9520.305^{+0.002}_{-0.002}$	&	$9520.306^{+0.002}_{-0.002}$	\\
$u_0$	&	$0.0360^{+0.0001}_{-0.0001}$	&	$-0.0360^{+0.0001}_{-0.0001}$	\\
$t_E$ [d]	&	$61.96^{+0.07}_{-0.05}$	&	$62.34^{+0.08}_{-0.07}$	\\
$\pi_{EN}$	&	$-0.1542^{+0.0168}_{-0.0169}$	&	$-0.1506^{+0.0170}_{-0.0162}$	\\
$\pi_{EE}$	&	$0.0107^{+0.0045}_{-0.0045}$	&	$-0.0114^{+0.0043}_{-0.0044}$	\\
$\rho_{lim}$	&	0.0092	&	0.0092	\\
G$_0$ [mag]	&	$15.7666^{+0.0002}_{-0.0002}$	&	$15.7666^{+0.0002}_{-0.0002}$	\\
$f_{s,G}$	&	$0.977^{+0.001}_{-0.001}$	&	$0.977^{+0.001}_{-0.001}$	\\
ATLAS o$_0$ [mag]	&	$15.5791^{+0.0004}_{-0.0004}$	&	$15.5792^{+0.0004}_{-0.0004}$	\\
$f_{s, ATLAS o}$	&	$0.968^{+0.001}_{-0.001}$	&	$0.968^{+0.001}_{-0.001}$	\\
ATLAS c$_0$ [mag]	&	$16.5658^{+0.0004}_{0.0004}$	&	$16.5658^{+0.0004}_{0.0004}$	\\
$f_{s,ATLAS c}$	&	$1.000^{+0.001}_{-0.001}$	&	$1.000^{+0.001}_{-0.001}$	\\
ZTF r$_0$ [mag]	&	$15.7368^{+0.0003}_{-0.0003}$	&	$15.7368^{+0.0003}_{-0.0003}$	\\
$f_{s, ZTF r}$	&	$0.902^{+0.001}_{-0.001}$	&	$0.902^{+0.001}_{-0.001}$	\\
ZTF g$_0$ [mag]	&	$17.3002^{+0.0002}_{-0.0002}$	&	$17.3002^{+0.0002}_{-0.0002}$	\\
$f_{s, ZTF g}$	&	$0.916^{+0.001}_{-0.001}$	&	$0.916^{+0.001}_{-0.001}$	\\
g$_0$ [mag]	&	$17.424^{+0.002}_{-0.002}$	&	$17.424^{+0.002}_{-0.002}$	\\
$f_{s,g}$	&	$0.954^{+0.001}_{-0.001}$	&	$0.954^{+0.001}_{-0.001}$	\\
i$_0$ [mag]	&	$15.199^{+0.002}_{-0.002}$	&	$15.199^{+0.002}_{-0.002}$	\\
$f_{s,i}$	&	$0.964^{+0.001}_{-0.002}$	&	$0.964^{+0.001}_{-0.001}$	\\
R$_0$ [mag]	&	$15.581^{+0.002}_{-0.002}$	&	$15.581^{+0.002}_{-0.002}$	\\
$f_{s,R}$	&	$0.953^{+0.002}_{-0.002}$	&	$0.954^{+0.002}_{-0.002}$	\\
I$_0$ [mag]	&	$14.631^{+0.002}_{-0.002}$	&	$14.631^{+0.002}_{-0.002}$	\\
$f_{s,I}$	&	$0.950^{+0.002}_{-0.002}$	&	$0.950^{+0.002}_{-0.002}$	\\
V$_0$ [mag]	&	$16.546^{+0.002}_{-0.002}$	&	$16.546^{+0.002}_{-0.002}$	\\
$f_{s,V}$	&	$0.934^{+0.002}_{-0.002}$	&	$0.934^{+0.002}_{-0.002}$	\\
B$_0$ [mag]	&	$18.127^{+0.002}_{-0.002}$	&	$18.127^{+0.002}_{-0.002}$	\\
$f_{s,B}$	&	$0.930^{+0.002}_{-0.002}$	&	$0.930^{+0.002}_{-0.002}$	\\
r$_0$ [mag]	&	$15.914^{+0.001}_{-0.001}$	&	$15.914^{+0.001}_{-0.001}$	\\
$f_{s,r}$	&	1.0 (fixed)	&	1.0 (fixed)	\\
$\chi^2$	&	5684	&	5685	\\
$\frac{\chi^2}{dof}$	&	0.92	&	0.92	\\
\hline

\end{tabular}
\end{scriptsize}
\end{table}

\FloatBarrier

\section{Acknowledgements} \label{app:acknowledgements}
The authors would like to thank Dr Radek Poleski for valuable discussions and comments during the development of this work. 

This work has made use of data from the European Space Agency (ESA) mission Gaia (\url{https://www.cosmos.esa.int/gaia}), processed by the Gaia
Data Processing and Analysis Consortium (DPAC,
\url{https://www.cosmos.esa.int/web/gaia/dpac/consortium}). Funding for the DPAC has been provided by national institutions, in particular the institutions
participating in the Gaia Multilateral Agreement. 

We acknowledge ESA Gaia, DPAC and the Photometric Science Alerts Team (\url{http://gsaweb.ast.cam.ac.uk/alerts}). 

BHTOM.space is based on the open-source TOM Toolkit by LCO and has been developed with funding from the OPTICON-RadioNet Pilot (ORP) of the European Union's Horizon 2020 research and innovation programme under grant agreement No 101004719 (2021-2025). This project has received funding from the European Union's Horizon Europe Research and Innovation programme ACME under grant agreement No 101131928 (2024-2028). 

Based on observations obtained with the Samuel Oschin Telescope 48-inch and the 60-inch Telescope at the Palomar
Observatory as part of the Zwicky Transient Facility project. ZTF is supported by the National Science Foundation under Grants
No. AST-1440341 and AST-2034437 and a collaboration including current partners Caltech, IPAC, the Oskar Klein Center at
Stockholm University, the University of Maryland, University of California, Berkeley , the University of Wisconsin at Milwaukee,
University of Warwick, Ruhr University, Cornell University, Northwestern University and Drexel University. Operations are
conducted by COO, IPAC, and UW.

This work has made use of data from the Asteroid Terrestrial-impact Last Alert System (ATLAS) project. The Asteroid Terrestrial-impact Last Alert System (ATLAS) project is primarily funded to search for near earth asteroids through NASA grants NN12AR55G, 80NSSC18K0284, and 80NSSC18K1575; byproducts of the NEO search include images and catalogs from the survey area. This work was partially funded by Kepler/K2 grant J1944/80NSSC19K0112 and HST GO-15889, and STFC grants ST/T000198/1 and ST/S006109/1. The ATLAS science products have been made possible through the contributions of the University of Hawaii Institute for Astronomy, the Queen’s University Belfast, the Space Telescope Science Institute, the South African Astronomical Observatory, and The Millennium Institute of Astrophysics (MAS), Chile.

The Pan-STARRS1 Surveys (PS1) and the PS1 public science archive have been made possible through contributions by the Institute for Astronomy, the University of Hawaii, the Pan-STARRS Project Office, the Max-Planck Society and its participating institutes, the Max Planck Institute for Astronomy, Heidelberg and the Max Planck Institute for Extraterrestrial Physics, Garching, The Johns Hopkins University, Durham University, the University of Edinburgh, the Queen's University Belfast, the Harvard-Smithsonian Center for Astrophysics, the Las Cumbres Observatory Global Telescope Network Incorporated, the National Central University of Taiwan, the Space Telescope Science Institute, the National Aeronautics and Space Administration under Grant No. NNX08AR22G issued through the Planetary Science Division of the NASA Science Mission Directorate, the National Science Foundation Grant No. AST-1238877, the University of Maryland, Eotvos Lorand University (ELTE), the Los Alamos National Laboratory, and the Gordon and Betty Moore Foundation.

Support for The Omega Key Project is provided by ANID's Millennium Science Initiative through grant ICN12\_009, awarded to the Millennium Institute of Astrophysics (MAS), and by ANID's Basal project FB210003.

YT acknowledges the support of DFG priority program SPP 1992 "Exploring the Diversity of Extrasolar Planets" (TS 356/3-1).
RAS and EB gratefully acknowledge support from the NASA XRP Program, through grant number 80NSSC19K0291.  

This work was authored by employees of Caltech/IPAC under Contract No. 80GSFC21R0032 with the National Aeronautics and Space Administration. 

The Omega Key Project has received funding from the European Union's Horizon 2020 research and innovation program under grant agreement No. 101004719 (OPTICON - RadioNet Pilot). 

This work is supported by the Polish MNiSW grant DIR/WK/2018/12. HHE thanks T\"urkiye National Observatories for partial support in using the TUG-T100 telescope with project number 21AT100-1799. 

Eda Sonbas thanks the Adiyaman University Astrophysics Application and Research Center for their support in the acquisition of data with the ADYU60 telescope. 

Josep Manel Carrasco work was (partially) supported by the Spanish MICIN/AEI/10.13039/501100011033 and by ''ERDF A way of making Europe'' by the ''European Union'' through grant PID2021-122842OB-C21, and the Institute of Cosmos Sciences University of Barcelona (ICCUB, Unidad de Excelencia ''Mar\'{\i}a de Maeztu'') through grant CEX2019-000918-M and the project 2021-SGR-00679 GRC de l'Ag\`encia de Gesti\'o d'Ajuts Universitaris i de Recerca (Generalitat de Catalunya). The Joan Oró Telescope (TJO) of the Montsec Observatory (OdM) is owned by the Catalan Government and operated by the Institute for Space Studies of Catalonia (IEEC).

G.D., M.S. and M.D.J acknowledge support by
the Astronomical Station Vidojevica and the Ministry of Science, Technological Development and Innovation of the Republic of Serbia (MSTDIRS) through contract no. 451-03-136/2025-03/200002 made with Astronomical Observatory (Belgrade), by the EC through project BELISSIMA (call FP7-REGPOT-2010-5, No. 256772), the observing and financial grant support from the Institute of Astronomy and Rozhen NAO BAS through the bilateral SANU-BAN joint research project "GAIA astrometry and fast variable astronomical objects", and support by the SANU project F-187. Also, this research was supported by the Science Fund of the Republic of Serbia, grant no. 6775, Urban Observatory of Belgrade - UrbObsBel. 

T.G. and the IST60 Telescope has been supported by the Scientific Research Projects Coordination Unit of Istanbul University (Project No.: TSG-2023-40046 and FBG-2017-23943) Turkish Republic, Directorate of Presidential Strategy and Budget project, 2016K121370. 

This work is partially supported by the Fundamental Fund of Thailand Science Research and Innovation (TSRI) through the National Astronomical Research Institute of Thailand (Public Organization) (FFB680072/0269). 

J.Z. and E.P. acknowledge funding from the Research Council of Lithuania (LMTLT, grant No. S-LL-24-1).

Some of the observations have been obtained with the 90cm Schmidt-Cassegrain Telescope (TSC90) in Piwnice Observatory, Institute of Astronomy of the Nicolaus Copernicus University in Toruń (Poland).

We acknowledge the use of ChatGPT-3.5 from OpenAI as a tool to improve the clarity and flow of the text. 

\end{appendix}

\end{document}